\documentclass[pra,superscriptaddress,twocolumn,showpacs,preprintnumbers,amsmath,amssymb]{revtex4-1}

\usepackage{graphicx}
\usepackage{dcolumn}
\usepackage{bm}
\usepackage{braket}
\usepackage{comment}

\usepackage{color} 				
\usepackage[normalem]{ulem} 			

\begin{document}

\title{Effects of anisotropy in simple lattice geometries on many-body properties of ultracold fermions in optical lattices}

\author{Anna Golubeva}
\affiliation{Institut f\"ur Theoretische Physik, Goethe-Universit\"at, 60438 Frankfurt/Main, Germany}

\author{Andrii Sotnikov}
\affiliation{Institut f\"ur Theoretische Physik, Goethe-Universit\"at, 60438 Frankfurt/Main, Germany}
\affiliation{Akhiezer Institute for Theoretical Physics, NSC KIPT, 61108 Kharkiv, Ukraine}

\author{Walter Hofstetter}
\affiliation{Institut f\"ur Theoretische Physik, Goethe-Universit\"at, 60438 Frankfurt/Main, Germany}

\date{\today}

\begin{abstract}
We study the effects of anisotropic hopping amplitudes on quantum phases of ultracold fermions in optical lattices described by the repulsive Fermi-Hubbard model. In particular, using dynamical mean-field theory (DMFT) we investigate the dimensional crossover between the isotropic square and the isotropic cubic lattice. We analyze the phase transition from the antiferromagnetic to the paramagnetic state and observe a significant change in the critical temperature: Depending on the interaction strength, the anisotropy can lead to both a suppression or increase. We also investigate the localization properties of the system, such as the compressibility and double occupancy. Using the local density approximation in combination with DMFT we conclude that density profiles can be used to detect the mentioned anisotropy-driven transitions.
\end{abstract}

\pacs{71.10.Fd, 75.50.Ee, 67.85.-d, 37.10.Jk}
\maketitle

\section{Introduction}
Ultracold atoms in optical lattices offer not only a clean and controlled quantum simulator for electronic solid-state materials, but also constitute interesting many-body systems in their own right, exhibiting a rich variety of physical phenomena. The high degree of tunability of the parameters allows to study the system in various regimes and thus to make direct quantitative comparisons with the predictions of theoretical models describing quantum many-body phenomena, such as the prominent Hubbard model~\cite{Hubbard1963}. A periodic potential is an essential ingredient of these models because it largely determines the electronic behavior in solid materials. Thus, the implementation of optical lattices in cold atom setups~\cite{Greiner2002Nat, Koehl2005} set a milestone in this experimental field. Since then, a substantial progress was made towards realizing the idea of a universal quantum simulator by preparing two-component mixtures of ultracold fermionic atoms in optical lattices. In particular, a fermionic Mott insulator was realized and detected~\cite{Joerdens2008Nat, Schneider2008Sci}, short-range antiferromagnetic correlations were observed and effectively measured~\cite{Greif2013S,Hart2015Nat}, and recently, the single-site resolution of fermions in optical lattices was experimentally attained \cite{Haller2015,Cheuk2015,Parsons2015}.

The lattice geometry is a crucial characteristic of these systems, since it essentially affects all the other physical properties. Experimentally, the geometry of an optical lattice is determined, among other things, by the spatial arrangement of the laser beams, and the possibility for the trapped atoms to tunnel within the lattice is then tuned via the laser intensity~\cite{Ess2010ARCMP, Blo2008RMP}. 
In particular, starting from an isotropic cubic lattice and gradually increasing the laser intensity along the $z$-direction results in a system consisting of separated layers of square lattices parallel to the $xy$-plane and thus effectively corresponds to a dimensional crossover from 3d to 2d (another type, from 3d to 1d, was studied in the related context in Refs.~\cite{Greif2013S, Imriska2014}). We analyze the effects of this dimensional crossover for a fermionic system with repulsive interactions. In particular, we study the magnetically ordered phases of the system and its thermodynamic properties. Employing a combination of DMFT and local density approximation (LDA+DMFT), we perform calculations for the case of an additional external potential (harmonic trap) and analyze the real-space particle distribution as well as the entropy of the system in different parameter regimes.

\section{System and Model}\label{sec.2}
The theoretical description is given by the Fermi-Hubbard model and the anisotropy is constituted in the ratio of the hopping parameters $t_z/t$, where the hopping parameters in the $xy$-plane are set equal ($t_x = t_y \equiv t$) and $t_z$ varies between zero and $t$ ($0 \leq t_z/t \leq 1$).
We consider a Fermi-Hubbard Hamiltonian of the following type:
\begin{eqnarray}\label{H}
\mathcal{\hat{H}}=&&
-\sum\limits_{\langle ij\rangle}\sum\limits_{\sigma} t_{ij} (\hat{c}^\dag_{i\sigma}\hat{c}_{j\sigma}+{\rm h.c.})
+U\sum\limits_{i}\hat{n}_{i\uparrow}\hat{n}_{i\downarrow}
\nonumber\\
&&+\sum\limits_{i}\sum\limits_{\sigma}(V_i-\mu)\hat{n}_{i\sigma},
\label{eq.1}
\end{eqnarray}
where the notation $\langle ij\rangle$ indicates a summation over nearest-neighbor sites. $t_{ij}$ is the hopping amplitude of the fermions with values for each spatial direction specified above and $U$ is the magnitude of the on-site repulsive ($U>0$) interaction of the two different species (or hyperfine states) $\sigma=\{\uparrow,\downarrow\}$.
The translational eigenstates in the periodic lattice potential $V_{lat}({\bf r}) =  \sum \limits_{\alpha=x,y,z} V_0^{(\alpha)} \cos^2(k \alpha)$ with lattice depth $V_0^{(\alpha)}$ can be represented in terms of Wannier orbitals. With $\psi({\bf r}-{\bf r}_i)$ being a single Wannier function localized at site $i$ the parameters for hopping and onsite interaction can be expressed in terms of overlap integrals:
\[
t_{ij} = - \int d^3r~\psi^*({\bf r}-{\bf r}_i) \left( -\frac{\hbar^2 \nabla^2}{2m} + V_{lat}({\bf r}) \right) \psi({\bf r}-{\bf r}_j),
\]
\[
U = \frac{4 \pi \hbar^2 a_s}{m} \int d^3r~ |\psi({\bf r})|^4.
\]
In expression~(\ref{H}), $\hat{c}^\dag_{i\sigma}$ ($\hat{c}_{i\sigma}$) is the corresponding creation (annihilation) operator of species with (pseudo)spin $\sigma$, with corresponding densities $\hat{n}_{i\uparrow}$ and $\hat{n}_{i\downarrow}$  ($\hat{n}_{i\sigma}=\hat{c}^\dag_{i\sigma}\hat{c}_{i\sigma}$). $V_i$ is the external (e.g., harmonic) potential at lattice site $i$, and $\mu$ is the chemical potential of the atoms. 
Note that we set the harmonic potential and the optical lattice to be independent of the atomic species.
The Hamiltonian~(\ref{eq.1}) implies a single-band approximation; in other words, we consider the case of a sufficiently strong lattice potential, $V_0 \gtrsim 5E_r$, with $E_r = \hbar^2k^2/(2m)$ being the recoil energy of the fermions.

\section{Method}\label{sec.3}
Our analysis is based on the dynamical mean-field theory (DMFT). In this approach the system is reduced to a local many-body problem described as a single lattice site (``impurity") coupled to an ``external bath", i.e. the spatial fluctuations are frozen, but the local dynamics of the system is fully preserved. The self-consistency conditions of DMFT can be derived in various ways. The approach introduced by A. Georges and G. Kotliar~\cite{Geo1996RMP} uses the Anderson impurity model~\cite{Anderson1961}, considering the single site as an ``impurity orbital'' and the bath as a ``conduction band'': 
\begin{eqnarray}\label{Eq:AM-H}
\mathcal{\hat{H}}_{AM} =&& - \sum \limits_{l, \sigma} \tilde \varepsilon_{l\sigma} \hat{a}_{l \sigma}^{\dagger} \hat{a}_{l \sigma} + \sum \limits_{l, \sigma} V_{l\sigma} (\hat{a}_{l \sigma}^{\dagger} \hat{c}_{\sigma} + \hat{c}_{\sigma}^{\dagger} \hat{a}_{l \sigma}) 
\nonumber\\
&&+ U \hat{n}_{\uparrow}^{c} \hat{n}_{\downarrow}^{c} - \mu (\hat{n}_{\downarrow}^{c} + \hat{n}_{\uparrow}^{c})
\end{eqnarray}
is the corresponding Hamiltonian with the fermionic creation and annihilation operators $\hat{c}_{\sigma}^{\dagger}, \hat{c}_{\sigma}$ describing the local degrees of freedom and $\hat{a}_{l \sigma}^{\dagger}, \hat{a}_{l \sigma}$ describing a set of non-interacting fermions representing the degrees of freedom of the effective bath acting on the site~$i$. Hence, the first term in the Hamiltonian~(\ref{Eq:AM-H}) describes the effective bath, the last two terms describe the impurity site, and the second term constitutes the coupling. Bath orbitals  are denoted by the index $l$, $\tilde \varepsilon_{l\sigma}$ and $V_{l\sigma}$ are the so-called Anderson parameters. Conveniently, the DMFT equations are expressed in the functional integral formalism. The key quantities in this notation are the so-called Weiss function $\mathcal{G}_{0,\sigma} (i \omega_{n})$ and the local propagator $G_{\sigma}(i \omega_{n})$, which are related to the self-energy $\Sigma_{\sigma}(i \omega_{n})$ by the Dyson equation:
\begin{eqnarray}\label{Eq:Dyson}
\Sigma_{\sigma}(i \omega_{n}) = \mathcal{G}_{0,\sigma}^{-1}(i \omega_{n}) - G_{\sigma}^{-1} (i \omega_{n}).
\end{eqnarray}
The Weiss function can be expressed as $\mathcal{G}_{0,\sigma}^{-1} (i \omega_{n}) = i \omega_{n} + \mu - \Delta_{\sigma} (i \omega_{n}),$ with the fermionic Matsubara frequency $\omega_n = (2n+1)\pi/\beta$, where $n$ is an integer number and $\beta$ is the inverse temperature, and the hybridization function $\Delta_{\sigma} (i \omega_{n})$, which is determined by the Anderson parameters,
\begin{eqnarray}\label{Eq:Hybr-f}
\Delta_{\sigma} (i \omega_{n}) = \sum \limits_{l,\sigma} \frac{|V_{l\sigma}|^2}{i \omega_{n} - \tilde \varepsilon_{l\sigma}}.
\end{eqnarray}
The self-consistency condition relating the dynamical mean-field $\Delta_{\sigma}(i \omega_{n})$ to the local Green's function $G_{\sigma}(i \omega_{n})$ is given by
\begin{align} \label{eq. SCC}
G_{\sigma} (i \omega_{n}) &= \int d\varepsilon \frac{D(\varepsilon)}{i \omega_{n} + \mu - \varepsilon - \Sigma_{\sigma}(i \omega_n)} \\
				 	 &= \int d\varepsilon \frac{D(\varepsilon)}{\Delta_{\sigma}(i \omega_{n}) + G^{-1}_{\sigma} (i \omega_{n}) - \varepsilon}.
\end{align}

{\it Impurity solver.} The Anderson impurity model can be solved by various techniques. We choose the exact diagonalization (ED) impurity solver \cite{Caffarel1994PRL}. Within this solver the number of orbitals in the effective Anderson model is considered to be finite. The algorithm essentially consists of three steps:
(i) The Weiss function is approximated by a discretized version. This replacement can be considered as a projection on a restricted functional subspace, which is the most involved part of the diagonalization procedure. The existing methods for the implementation of this projection are described in reference~\cite{Geo1996RMP}. (ii) The Anderson Hamiltonian~(\ref{Eq:AM-H}) restricted to a finite number ($n_s$) of orbitals is diagonalized exactly and the local propagator $G_\sigma(i\omega_n)$ is computed. (iii) The self-consistency condition yields a new Weiss function $\mathcal{G}_{0,\sigma}$, which in turn is approximated by a function $\mathcal{G}_{0,\sigma}^{(n_s)}$ with a new set of Anderson parameters $\tilde \varepsilon_{l\sigma}$ and $V_{l\sigma}$. The process is iterated until a converged set of parameters is reached. The full numerical diagonalization is currently feasible up to $n_s = 7$. Our results are based on calculations with $n_s=5$.

{\it Two-sublattice DMFT.}
Bipartite structures, such as states with antiferromagnetic order, can be described in an appropriate way by introducing two sublattices. 
Within the DMFT approach one then needs to solve the impurity problem twice on two adjacent sites of the original lattice. The self-consistency condition for this case is given by
\begin{equation} \label{eq.12}
 G_{\sigma}^{\alpha}(i\omega_n) = \zeta_{\sigma}^{\bar{\alpha}}
 \int d\varepsilon \frac{D(\varepsilon)}{\zeta_{\sigma}^A \zeta_{\sigma}^B-\varepsilon^2}
\end{equation}
with $\zeta_{\sigma}^{\bar{\alpha}} = i\omega_n + \mu - \Sigma_{\sigma}^{\bar{\alpha}}$, and the sublattice indices $\alpha = A, B$ and its opposite $\bar{\alpha} = B, A$.
In the balanced case (i.e. no imbalance in the hopping amplitudes or chemical potentials between two spin components \cite{Sotnikov2013}), there is an additional symmetry present, $\zeta_{\sigma}^A=\zeta_{\bar{\sigma}}^B$, which allows to reduce the computations related to the solution of the impurity problem to a single site.

Note that (apart from the bipartite structure of the self-consistency condition) the geometry of the lattice enters our numerical studies only through the non-interacting density of states $D(\varepsilon)$. As can be seen in Fig.~\ref{fig1}, this quantity depends strongly on the possible anisotropies in the system. In Appendix~\ref{app.A} we derive analytical expressions for $D(\varepsilon)$. The evaluation of elliptic integrals involved in these expressions was done by the AGM method and is presented in Appendix~\ref{app.B}.

\begin{figure}
\includegraphics[width=0.5\textwidth]{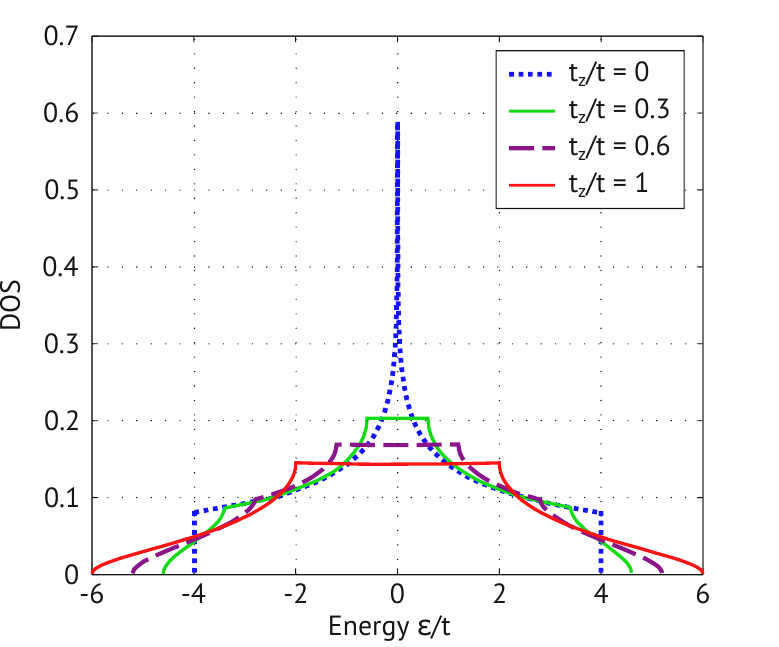}
\caption{(Color online) Non-interacting density of states profiles for $t_x/t=t_y/t=1$ and different $t_z$-values.}
\label{fig1}
\end{figure}

\section{Results}\label{sec.4}
In this section we discuss how the introduced spatial anisotropy in hopping amplitudes influences the main many-body properties of the system.

Preceding the following discussion it should be noted that DMFT calculations for a 2d system predict antiferromagnetic long-range order (LRO), whereas it is well known that according to the Mermin-Wagner-Hohenberg theorem LRO can arise in the isotropic (SU(2) symmetric) Hubbard model at finite temperatures only for $d > 2$~\cite{MerminWagner1966}. Thus, in low-dimensional systems ($d \leq 2$) LRO for this model is only possible at $T=0$. However, the DMFT results for these systems do have a physical relevance -- the calculated N\'{e}el temperature gives a good quantitative estimate for the boundary at which (short-ranged) antiferromagnetic correlations arise in the system~\cite{Gorelik2011JLTP}.

\subsection{Magnetically ordered phases}
\begin{figure}
\includegraphics[width=0.5\textwidth]{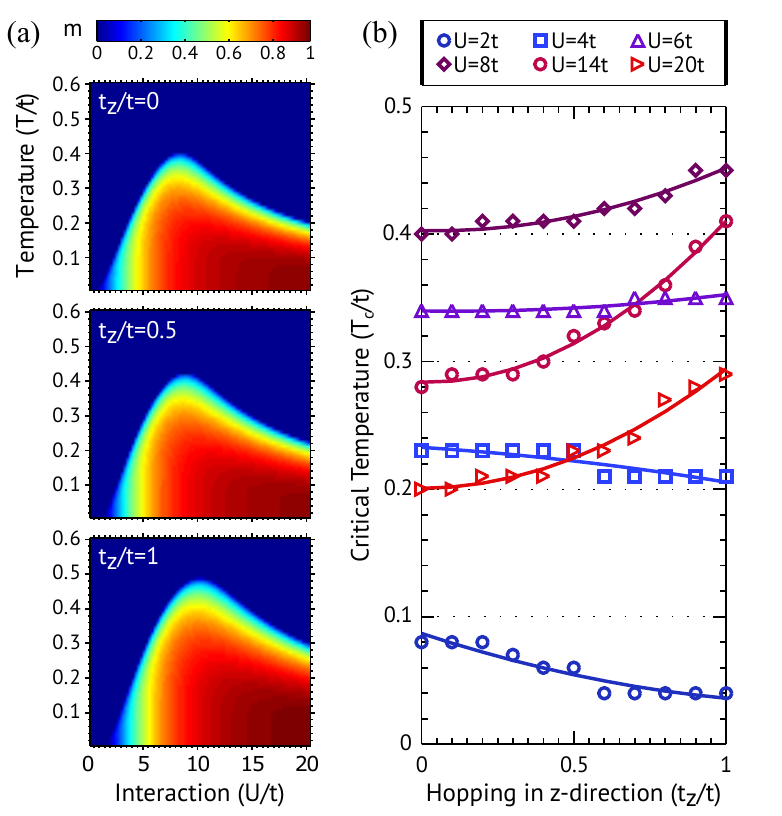}
\caption{(Color online) (a) Phase diagrams for the N\'{e}el-ordered phases in lattices with the hopping parameter $t_{z}/t\,=\,0,\,0.5,\,1$ obtained by DMFT at half filling ($\mu = U/2$). The antiferromagnetic order parameter, staggered magnetization $m = |n_{i\uparrow}-n_{i\downarrow}|$, is color-coded in the $T/U$-plane. (b) Dependence of the critical temperature $T_{c}$ on the hopping parameter $t_z$ for different interaction strengths $U$. The critical temperature is nearly constant for $U = 6t$ and it increases (decreases) for larger (smaller) values of $U$.}
\label{fig2}
\end{figure}

As expected on the basis of known limiting cases for isotropic hopping in square ($t_{z}/t = 0$) and cubic ($t_{z}/t=1$) lattice geometries, we observe a continuous change of the ordered phase region with an increase of $t_{z}$ as shown in Fig.~\ref{fig2}(a). In particular, at strong coupling ($U \gtrsim 5t$) the critical temperature increases significantly, in the region of the intermediate interaction strength $U/t$ the phase transition line stays approximately unchanged, but at low $U/t$ the antiferromagnetic phase is more suppressed. To demonstrate these effects we plot the critical temperature as a function of $t_z$ for different interaction strengths (Fig.~\ref{fig2}(b)). With increasing values of $t_z$, the critical temperature decreases for weak interactions, rises for strong interactions and stays approximately unchanged for $U / t = 6$. This behavior becomes plausible when we consider the known analytical results for the limiting cases.

In the case $U \ll Zt$ the critical temperature is given by
\begin{displaymath}
T_{c} \propto \sqrt{Z} t e^{-c \frac{\sqrt{Z}t}{U}} \end{displaymath}
with the coordination number $Z$ and a positive constant $c$ \cite{vanDongen}. In this regime, the exponential suppression dominates over the linear growth with increase of the coordination number $Z$ which corresponds to increase of $t_{z}$ in our case.
For the opposite case of strong coupling $U \gg Zt$ the Heisenberg spin model is applicable. Within the mean-field description  $T_{c}$ is proportional to the magnetic coupling constant $J$,
\begin{displaymath}
T_{c} \propto J=\frac{Zt^{2}}{U}. \end{displaymath}
As one can see from Fig.~\ref{fig2}(b), the mean-field relation between the critical temperatures in simple cubic and square lattices, that according to the last formula is given by $T_{c}^{\text{(3d)}}/T_{c}^{\text{(2d)}}= 3/2$, is approximately fulfilled for $U~\gtrsim~20~t$.

\subsection{Compressibility and double occupancy}
\begin{figure}
\includegraphics[width=0.35\textwidth]{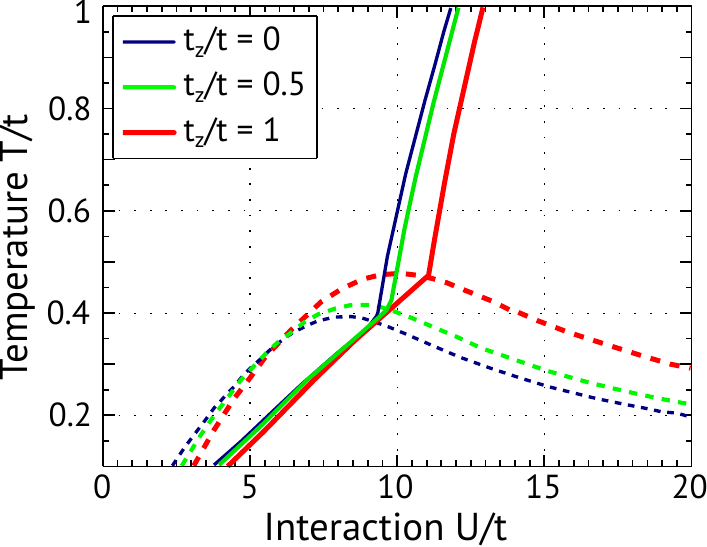}
\caption{(Color online) Contour lines ($\kappa_e = 0.01/t$) (solid) and magnetic transition lines (dashed) for $t_z / t = 0, 0.5, 1$. For larger $t_z$ the crossover to the insulating regime is shifted towards larger interaction strength values.}
\label{fig3}
\end{figure}

The electronic conductivity is an important characteristic of solid materials. However, in systems of ultracold atoms in optical lattices, it is easier to measure the compressibility, which is qualitatively related to the conductivity: an electronic system is conducting if the many-body state of the electrons is compressible. This electronic compressibility is defined as the derivative of the electron density with respect to the chemical potential, $\kappa_e \equiv \frac{\partial n}{\partial \mu}$. Fig.~\ref{fig3} shows the comparison of the contour lines with $\kappa_e = 0.01/t$ for particular values of $t_z$. The shift of the contour line towards larger interaction strengths for larger $t_z$ in the unordered phase shows that at certain parameter values of $U$ and $T$ it is experimentally possible to change the compressibility of the system in a wide range by just varying the hopping parameter $t_z$ (i.e. changing the laser intensity in one direction) and keeping $U$ and $T$ constant.

\begin{figure}
\includegraphics[width=0.5\textwidth]{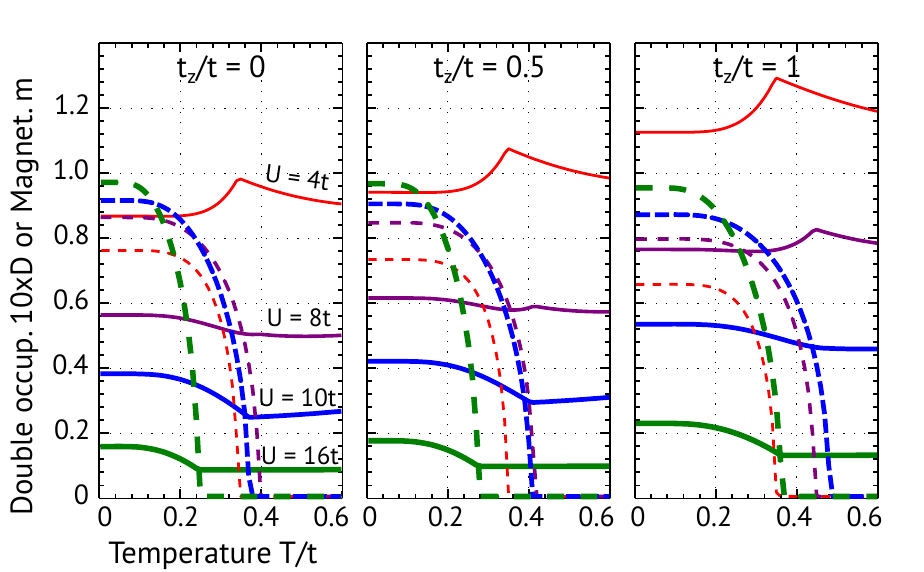}
\caption{(Color online) Double occupancy (solid lines) and magnetization (dashed lines) plotted as a function of temperature for different values of the interaction strength $U$ and $t_z/t = 0, 0.5, 1$. At large interaction strengths a pronounced increase of double occupancy is observed below the N\'{e}el temperature.}
\label{fig4}
\end{figure}

The average on-site double occupancy given as $D_i = \langle \hat{n}_{i \uparrow} \hat{n}_{i \downarrow} \rangle$ is another useful indicator for the degree of localization of the fermions and an observable quantity of great importance in optical lattice experiments~\cite{Joerdens2008Nat, Schneider2008Sci, Scarola2009}. In the regime of strong coupling and low temperatures it was found to exhibit an increase in the presence of antiferromagnetic correlations that appear with decreasing temperature~\cite{Gor2010PRL, Khatami2011}.
In order to analyze the relationship between the double occupancy and magnetic ordering, we plot both $D$ and $m$ as a function of temperature in one diagram (Fig.~\ref{fig4}). We clearly see that the onset of staggered magnetization coincides with an enhancement in double occupancy at large interaction strengths, in accordance with the results given in~\cite{Gor2010PRL}. We also find that at given system parameters for the two-dimensional case ($t_z/t = 0$) there is a good agreement with the results obtained by means of numerical linked cluster expansions technique in~\cite{Khatami2011}. In the three-dimensional case, however, our observations show stronger dependencies in comparison with more accurate dynamical cluster approximation results~\cite{Fuchs2011}. At intermediate coupling, the signal becomes less pronounced, and below a certain value of $U$, we even observe converse behavior: the double occupancy smoothly increases just before the system enters the antiferromagnetic phase and then decreases with increasing magnetization.
Comparing these diagrams for different $t_z$ we note that the effect of the double occupancy enhancement due to antiferromagnetic correlations occurs at larger $U$ for larger $t_z$ values, e.g. at $U \approx10t$ for $t_z = t$ and at $U \approx 8t$ for $t_z = 0$. These values correspond to the paramagnetic crossover region between the metallic and Mott insulating states, which is shifted to lower interaction strengths for smaller $t_z$-values (see also Fig.~\ref{fig3}).

In general, we observe higher double occupancy at lower interaction strength. This effect is directly understandable from the model Hamiltonian (\ref{H}): double occupation of a particular lattice site becomes energetically unfavorable in case of high interaction ``costs". The increase of double occupancy in the antiferromagnetic phase at strong coupling is an effect of virtual hopping: In the antiferromagnetically ordered system each particle is surrounded by particles with an opposite spin, allowing it to virtually hop to all $Z$ next neighbors and by this to lower its energy, whereas in the paramagnetic regime only 50\% of the next neighbors have the opposite spin (see also Ref.~\cite{Gor2010PRL} for details). In the regime of weak coupling, the paramagnetic state is metallic, i.e. the particles are itinerant. Thus, the double occupancy is generally large. However, the temperature affects the metallic quality of the state: at higher temperatures, the delocalization of particles is disturbed by thermal fluctuations. Therefore, the delocalization and, respectively, the double occupancy increases at lower temperature values. The peculiar cusps in the curve shape emerge at the transition to the antiferromagnetic phase: the double occupancy decreases because the antiferromagnetic ordering opens a gap in the charge excitations spectrum, i.e. the system becomes an insulator.

\begin{figure}
\includegraphics[width=0.5\textwidth]{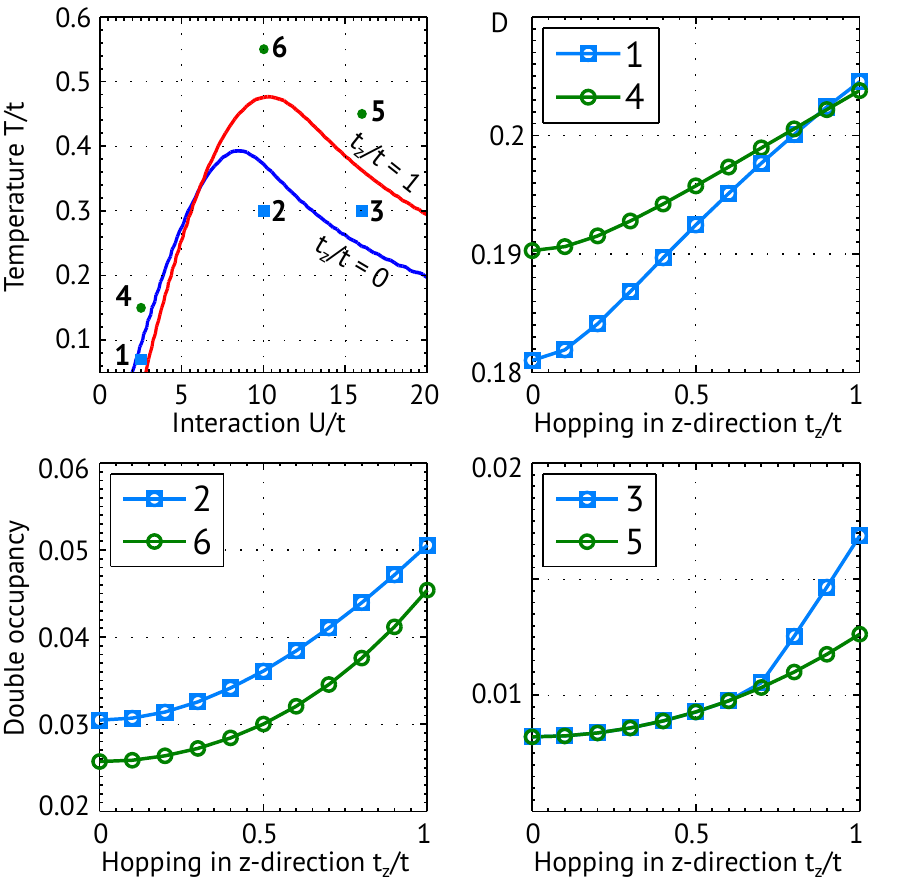}
\caption{(Color online) Points in the magnetic phase diagram with transition lines between the PM and AFM phase for 2d ($t_z/t = 0$) and 3d ($t_z/t = 1$) and analysis of the pairs of points from adjacent regions. Double occupancy is plotted against $t_z$. For each curve, the values for $T$ and $U$ are kept constant. (1): $U = 2.5 t, T = 0.07 t$, (2): $U = 10 t, T = 0.2 t$, (3): $U = 15 t, T = 0.35 t$, (4): $U = 2.5 t, T = 0.25 t$, (5): $U = 17.5 t, T = 0.4 t$, (6): $U = 10 t, T = 0.55 t$.}
\label{fig5}
\end{figure}

To analyze how the system's dimensionality affects the double occupancy we chose particular pairs of points in the magnetic phase diagram, which are indicated in Fig.~\ref{fig5}, and found a pronounced dependence of $D$ on the value of $t_z$. The increase of $D$ is particularly strong when the change in $t_z$ from zero to $t$ induces a phase transition, as it is the case in the points 1 (AFM to PM) and 3 (PM to AFM). In the points 2 and 4-6 the phase does not change for any value of $t_z\in[0,t]$, thus the curves have nearly the same slopes. The resulting dependence of the double occupancy on the hopping parameter $t_z$ can be used to identify the presence of magnetic correlations or to measure the temperature of the system, since $D$ is an experimentally accessible observable, $t_z$ is a tunable parameter, and the interaction strength $U$ is a known quantity of the experimental setup. However, it was pointed out in~\cite{Khatami2011} that the increase of double occupancy in the considered parameter regime occurs not only in the Mott-insulating core, but also in areas with $n < 1$. Thus, to identify the increase of $D$ as a signal of onsetting antiferromagnetic order in experiments, it has to be additionally ensured that the density in the measured region of the trap is close to $n = 1$.

\subsection{Entropy analysis}
\begin{figure}
\includegraphics[width=0.5\textwidth]{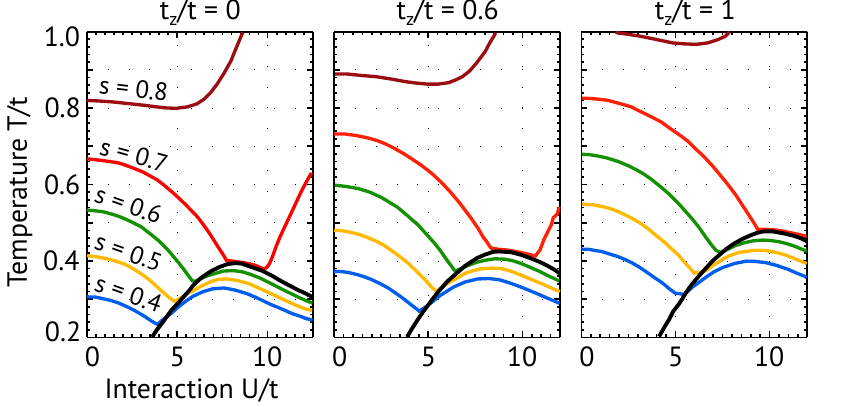}
\caption{(Color online). Isentropic curves in the $T/U$-plane obtained by LDA+DMFT at half filling ($\mu_0~=~U/2$) for the hopping parameters $t_z/t=0, 0.6, 1$.}
\label{fig6}
\end{figure}

In experiments with ultracold atoms the system is characterized in a natural way by the total particle number $N$ and the total entropy $S$. These quantities can ideally be kept fixed, since the system is isolated from the environment and the experiments are assumed to be carried out adiabatically. The entropy per site, however, is variable and is often used as a thermometer, since it is independent of the trap frequency, while the temperature changes even when the trap frequency is increased adiabatically~\cite{SchneiderThesis}.
To determine the entropy theoretically, we use the definition from statistical thermodynamics: $S = \ln{\Omega}$ ($k_B = 1$), where $\Omega$ is the number of possible microstates of the system under consideration.
In the Hubbard model within the single-band approximation (\ref{eq.1}), a single site can be found in four different states, thus the maximal entropy per site is given by $s_{\max} = \ln(4)$. This limit is reached at half filling (the number of particles coincides with the number of sites) and vanishing interaction $U=0$, where all four possible states have the same probability in the framework of the microcanonical description. In case of strong interactions, however, the system becomes a Mott insulator with only two possible states ($\ket{\uparrow}$ and $\ket{\downarrow}$), such that the upper entropy bound is given by $s = \ln(2) \approx 0.69$.
The calculations with the combined DMFT+LDA method provide us with expectation values of the local particle density (``filling'') $n$ at different values of the chemical potential. Given these quantities, we can use a Maxwell relation for the entropy per site~\cite{Sotnikov2013}:
\begin{eqnarray}
\frac{\partial s}{\partial \mu} = \frac{\partial n}{\partial T} ~~~ \longrightarrow ~~~ s(\mu_0, T) = \int \limits_{- \infty}^{\mu_0} \left( \frac{\partial n}{\partial T} \right) d \mu.
\end{eqnarray}
Fig.~\ref{fig6} shows the isentropic curves in the $T/U$-plane obtained for different values of the hopping parameter~$t_z$. Naturally, larger entropy values are found at higher temperature. We observe a characteristic shape of the isentropic curves for small entropy values ($s\leq0.7$), which have a negative slope at weak coupling, bend sharply when entering the antiferromagnetic region and then follow the curvature of the magnetic phase transition boundary (black curve). On the left side of the phase diagram the so-called Pomeranchuk cooling effect can be observed: an increase of the interaction strength $U$ at constant entropy leads to a decrease in temperature beyond non-interacting band structure effects~\cite{Wer2005PRL}. The isentropic curves are shifted towards larger values of $T$ and $U$ when $t_z$ increases, which makes the region with the Pomeranchuk effect larger. However, according to an additional analysis performed we can conclude that this enlargement is mainly based on the increase of the bandwidth.
Comparing our results in the limiting cases $t_z/t = 0$ and $t_z/t = 1$ to references~\cite{Khatami2011} and~\cite{Kozik2013, Fuchs2011}, respectively, we observe nearly the same quantitative overestimates of critical entropies by the DMFT method. At the same time, the qualitative behavior persists and agrees well in both limits, and therefore we expect that the results in the dimensional crossover region remain of a high importance.

\begin{figure}
\includegraphics[width=0.5\textwidth]{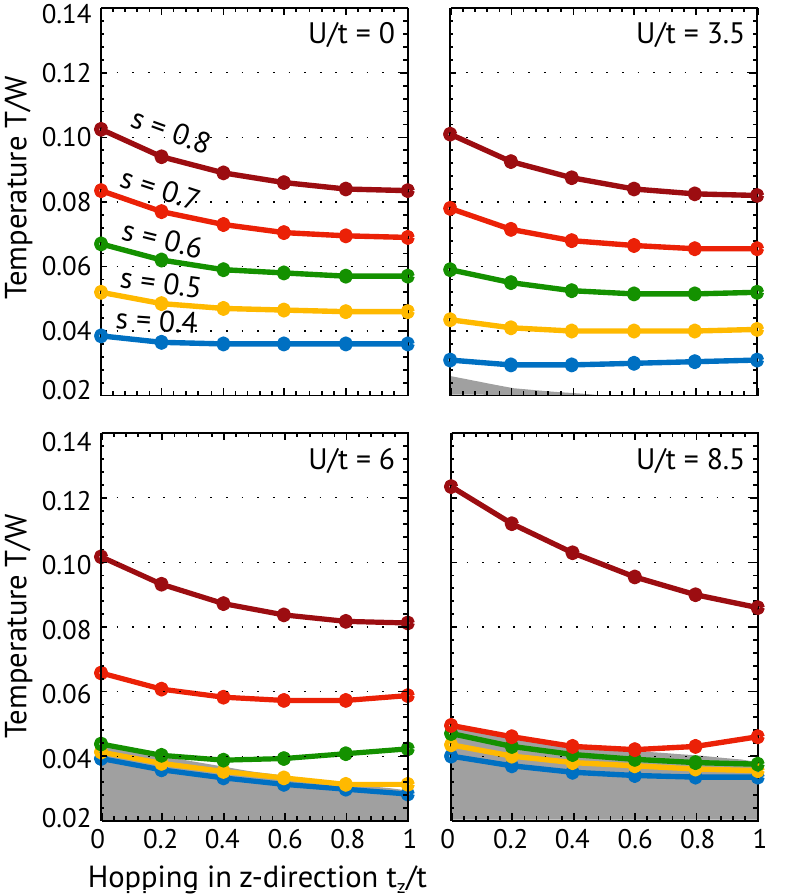}
\caption{(Color online). Temperature $T$, scaled by the bandwidth $W=8t+4t_z$, is plotted as a function of the hopping parameter $t_z/t$ for fixed values of entropy $s$ at different interaction strengths $U/t = 0, 3.5, 6, 8.5$.}
\label{fig7}
\end{figure}

To analyze the effects of hopping anisotropy, we plot the temperature, scaled by the bandwidth $W$ (in order to exclude its broadening effect), as a function of the hopping parameter~$t_z$ for particular entropy values. Fig.~\ref{fig7} shows these plots for four different interaction strengths. We observe that, in general, in the 2d-case ($t_z/t = 0$) the rescaled temperature at a given entropy value is larger than in the 3d-case ($t_z/t = 1$). This effect can be easily understood when we consider the entropy as a measure for disorder: the transition from 2d into 3d increases the number of the nearest neighbors from $Z=4$ to $Z=6$, opening two more tunneling options for a single site (i.e. the positive and the negative $z$-direction). Hence, the entropy of the system should increase, but since its value is kept fixed, its conjugated thermodynamic variable -- the temperature -- decreases.
However, the decrease of $T/W$ is not monotone in general: for particular $U$- and $s$-values, the isentropic curves exhibit a minimum at intermediate $t_z$-values. This effect is based on the magnetic phase transitions of the system. The slope of the isentropic curves becomes positive when the initially antiferromagnetically ordered system (regions shaded in grey in the figure) turns into the paramagnetic state. Again, this effect can be explained in the entropy picture: In the antiferromagnetic phase the spins are ordered in a checkerboard pattern, so that hopping is possible to each of the nearest neighbor sites. In the paramagnetic phase, however, on average only 50\% of the nearest neighbors have an opposite spin. Thus, with the transition into the paramagnetic state the entropy should decrease, and by the same argument as above the temperature increases.

\subsection{Signatures of anisotropy-driven transitions in the real-space density profiles}

\begin{figure}
\includegraphics[width=0.5\textwidth]{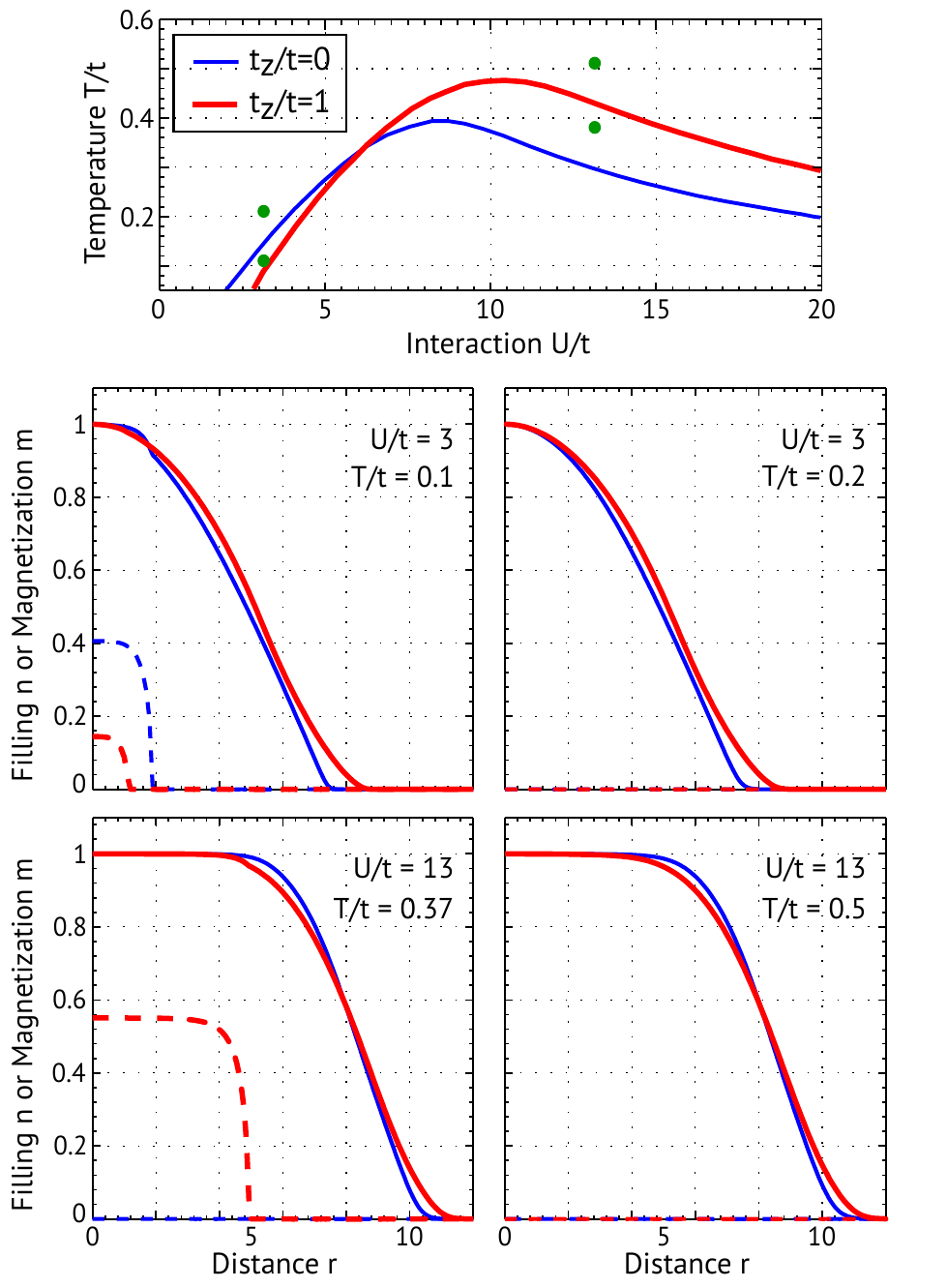}
\caption{(Color online). A selection of points in the magnetic phase diagram for the analysis of the particle density distribution in a harmonic trap and its dependence on the quantum phase in the system. Below: Particle distribution in real space (solid line) and magnetization (dashed line) for $t_z/t = 0$ (blue, thin) and $t_z/t = 1$ (red, thick), in the regimes of weak ($U=3t$) and strong ($U=13t$) coupling at different temperatures. On the $x$-axis is the radial distance from the trap center, measured in units of the lattice constant $a=1$. The trapping potential is $V=0.1 t$.}
\label{fig8}
\end{figure}

The real space density distribution of the atoms is experimentally accessible with in-situ imaging techniques~\cite{Bakr2009Nat, Sherson2010, Parsons2015, Cheuk2015, Haller2015}. Within the local density approximation this is equivalent to measuring the density as a function of chemical potential for a given temperature~\cite{SchneiderThesis}. We use the data obtained by the LDA+DMFT calculations performed for different parameter regimes, which are indicated by points in the upper diagram in Fig.~\ref{fig8}. The resulting particle density profiles and the magnetization profiles in the trap are shown in below. 
Note that at lower temperature the 2d- and the 3d-system are in different magnetic phases: In the regime of weak coupling $U=3t$ at $T=0.1t$ the 2d-system is in the antiferromagnetic insulating state, whereas the 3d-system is in the region of the paramagnetic metal (although very close to the magnetic transition line, thus a finite magnetization is present). Similarly, in the regime of strong interaction $U=13t$ at $T=0.37t$ the 2d-case corresponds to the paramagnetic insulator, while the 3d-case is in the antiferromagnetic insulating phase.
In each case the particle density equals one in the trap center and decreases with increasing distance from the trap center to finally vanish at certain values of $r$. The shape of the density distribution differs depending on the parameter values $U$ and $T$. In the regime of strong coupling ($U/t = 13$) both density profiles exhibit a wide plateau in the trap center, indicating an insulating state. However, in case of weak interaction this plateau can be identified as a signature of the antiferromagnetically ordered state: it only appears when significant staggered magnetization is present. Thus, in the regime of weak coupling the real-space density profile can be used in the experiment to detect an antiferromagnetic phase without measuring local quantities. However, in other regimes it remains essential to access local observables such as magnetization and double occupancy in order to determine the system's phase.
The sharp kinks, observable in case of $U/t = 3, T/t = 0.1$ (2d-curve) and $U/t = 13, T/t = 0.37$ (3d-curve), which seem to indicate the appearance of the antiferromagnetic phase, are known to be artifacts of LDA+DMFT and are smoothed out in the real system by proximity effects.

\section{Conclusions}
We studied the effects of hopping anisotropy on the physical properties of fermionic atoms in optical lattices with simple cubic geometry. The analysis is based on calculations for the hopping parameters $t_{x} = t_{y} \equiv t$ and $t_{z}/t \in [0.0, 1.0]$ performed with the DMFT approach. We found that all characteristic many-body properties and quantum phases of the system depend on $t_z$. In particular, we analyzed the significant changes in the magnetic phase diagram that depend on the interaction strength in a non-trivial way: the critical temperature for antiferromagnetic long-range order decreases with $t_z$ at small values of $U$, but increases with $t_z$ in the regime of strong coupling. This behavior is consistent with the analytical expressions from the mean-field approaches for the limiting cases $U \gg Zt$ and $U \ll Zt$, which have different dependence on the coordination number $Z$.

The analysis of metallic vs. insulating properties showed that it is experimentally possible to change the compressibility of the system by just varying $t_z$ and keeping the other parameters ($U$ and $T$) constant. The double occupancy increases with increasing $t_z$ and shows a pronounced dependence on the magnetic phase of the system. Thus, in the experiments the double occupancy can be used to determine the presence of magnetic correlations in the system or to measure the temperature. However, we note that the obtained results for the double occupancy are based on DMFT, i.e., momentum-independent analysis. Thus, the influence of non-local correlations close to transitions is not properly accouted for and this can lead to some deviations between theoretical predictions and experimental results. Therefore, a further improvement in this direction would be a theoretical analysis of the double occupancy by more accurate approach, e.g. the dynamical cluster approximation~\cite{Imriska2014} or diagrammatic determinant Monte-Carlo method~\cite{Kozik2013}, in the region of the studied dimensional crossover.

In combination with the local density approximation (DMFT+LDA), we analyzed the entropy and the real-space density profiles of the system with a harmonic trapping potential.
In the analysis of the entropy we find that the parameter region with the presence of the Pomeranchuk effect becomes larger with increasing $t_z$. At constant entropy, the rescaled temperature generally decreases with $t_z$, which is simply explained by the increase of the coordination number. However, when the initially antiferromagnetically ordered system turns into the paramagnetic state, the temperature rises again.
Our analysis of the particle density distribution in real-space showed that at weak coupling the transition into the magnetically ordered state corresponds to the appearance of a plateau in the density profile. Thus, this signal could also be used to detect the antiferromagnetic phase in the experiment.

\begin{acknowledgments}
Support by the Deutsche Forschungsgemeinschaft DFG via Sonderforschungsbereich SFB/TR 49 and Forschergruppe FOR 801 is gratefully acknowledged.
\end{acknowledgments}

\bibliography{A17}

\begin{thebibliography}{29}%
\makeatletter
\providecommand \@ifxundefined [1]{%
 \@ifx{#1\undefined}
}%
\providecommand \@ifnum [1]{%
 \ifnum #1\expandafter \@firstoftwo
 \else \expandafter \@secondoftwo
 \fi
}%
\providecommand \@ifx [1]{%
 \ifx #1\expandafter \@firstoftwo
 \else \expandafter \@secondoftwo
 \fi
}%
\providecommand \natexlab [1]{#1}%
\providecommand \enquote  [1]{``#1''}%
\providecommand \bibnamefont  [1]{#1}%
\providecommand \bibfnamefont [1]{#1}%
\providecommand \citenamefont [1]{#1}%
\providecommand \href@noop [0]{\@secondoftwo}%
\providecommand \href [0]{\begingroup \@sanitize@url \@href}%
\providecommand \@href[1]{\@@startlink{#1}\@@href}%
\providecommand \@@href[1]{\endgroup#1\@@endlink}%
\providecommand \@sanitize@url [0]{\catcode `\\12\catcode `\$12\catcode
  `\&12\catcode `\#12\catcode `\^12\catcode `\_12\catcode `\%12\relax}%
\providecommand \@@startlink[1]{}%
\providecommand \@@endlink[0]{}%
\providecommand \url  [0]{\begingroup\@sanitize@url \@url }%
\providecommand \@url [1]{\endgroup\@href {#1}{\urlprefix }}%
\providecommand \urlprefix  [0]{URL }%
\providecommand \Eprint [0]{\href }%
\providecommand \doibase [0]{http://dx.doi.org/}%
\providecommand \selectlanguage [0]{\@gobble}%
\providecommand \bibinfo  [0]{\@secondoftwo}%
\providecommand \bibfield  [0]{\@secondoftwo}%
\providecommand \translation [1]{[#1]}%
\providecommand \BibitemOpen [0]{}%
\providecommand \bibitemStop [0]{}%
\providecommand \bibitemNoStop [0]{.\EOS\space}%
\providecommand \EOS [0]{\spacefactor3000\relax}%
\providecommand \BibitemShut  [1]{\csname bibitem#1\endcsname}%
\let\auto@bib@innerbib\@empty
\bibitem [{\citenamefont {Hubbard}(1963)}]{Hubbard1963}%
  \BibitemOpen
  \bibfield  {author} {\bibinfo {author} {\bibfnamefont {J.}~\bibnamefont
  {Hubbard}},\ }\href {\doibase 10.1098/rspa.1963.0204} {\bibfield  {journal}
  {\bibinfo  {journal} {Proceedings of the Royal Society of London A:
  Mathematical, Physical and Engineering Sciences}\ }\textbf {\bibinfo {volume}
  {276}},\ \bibinfo {pages} {238} (\bibinfo {year} {1963})}\BibitemShut
  {NoStop}%
\bibitem [{\citenamefont {Greiner}\ \emph {et~al.}(2002)\citenamefont
  {Greiner}, \citenamefont {Mandel}, \citenamefont {Esslinger}, \citenamefont
  {H\"ansch},\ and\ \citenamefont {Bloch}}]{Greiner2002Nat}%
  \BibitemOpen
  \bibfield  {author} {\bibinfo {author} {\bibfnamefont {M.}~\bibnamefont
  {Greiner}}, \bibinfo {author} {\bibfnamefont {O.}~\bibnamefont {Mandel}},
  \bibinfo {author} {\bibfnamefont {T.}~\bibnamefont {Esslinger}}, \bibinfo
  {author} {\bibfnamefont {T.~W.}\ \bibnamefont {H\"ansch}}, \ and\ \bibinfo
  {author} {\bibfnamefont {I.}~\bibnamefont {Bloch}},\ }\href
  {http://dx.doi.org/10.1038/415039a} {\bibfield  {journal} {\bibinfo
  {journal} {Nature}\ }\textbf {\bibinfo {volume} {415}},\ \bibinfo {pages}
  {39} (\bibinfo {year} {2002})}\BibitemShut {NoStop}%
\bibitem [{\citenamefont {K\"ohl}\ \emph {et~al.}(2005)\citenamefont {K\"ohl},
  \citenamefont {Moritz}, \citenamefont {St\"oferle}, \citenamefont
  {G\"unter},\ and\ \citenamefont {Esslinger}}]{Koehl2005}%
  \BibitemOpen
  \bibfield  {author} {\bibinfo {author} {\bibfnamefont {M.}~\bibnamefont
  {K\"ohl}}, \bibinfo {author} {\bibfnamefont {H.}~\bibnamefont {Moritz}},
  \bibinfo {author} {\bibfnamefont {T.}~\bibnamefont {St\"oferle}}, \bibinfo
  {author} {\bibfnamefont {K.}~\bibnamefont {G\"unter}}, \ and\ \bibinfo
  {author} {\bibfnamefont {T.}~\bibnamefont {Esslinger}},\ }\href {\doibase
  10.1103/PhysRevLett.94.080403} {\bibfield  {journal} {\bibinfo  {journal}
  {Phys. Rev. Lett.}\ }\textbf {\bibinfo {volume} {94}},\ \bibinfo {pages}
  {080403} (\bibinfo {year} {2005})}\BibitemShut {NoStop}%
\bibitem [{\citenamefont {J\"ordens}\ \emph {et~al.}(2008)\citenamefont
  {J\"ordens}, \citenamefont {Strohmaier}, \citenamefont {Gunter},
  \citenamefont {Moritz},\ and\ \citenamefont {Esslinger}}]{Joerdens2008Nat}%
  \BibitemOpen
  \bibfield  {author} {\bibinfo {author} {\bibfnamefont {R.}~\bibnamefont
  {J\"ordens}}, \bibinfo {author} {\bibfnamefont {N.}~\bibnamefont
  {Strohmaier}}, \bibinfo {author} {\bibfnamefont {K.}~\bibnamefont {Gunter}},
  \bibinfo {author} {\bibfnamefont {H.}~\bibnamefont {Moritz}}, \ and\ \bibinfo
  {author} {\bibfnamefont {T.}~\bibnamefont {Esslinger}},\ }\href
  {http://dx.doi.org/10.1038/nature07244} {\bibfield  {journal} {\bibinfo
  {journal} {Nature}\ }\textbf {\bibinfo {volume} {455}},\ \bibinfo {pages}
  {204} (\bibinfo {year} {2008})}\BibitemShut {NoStop}%
\bibitem [{\citenamefont {Schneider}\ \emph {et~al.}(2008)\citenamefont
  {Schneider}, \citenamefont {Hackerm\"uller}, \citenamefont {Will},
  \citenamefont {Best}, \citenamefont {Bloch}, \citenamefont {Costi},
  \citenamefont {Helmes}, \citenamefont {Rasch},\ and\ \citenamefont
  {Rosch}}]{Schneider2008Sci}%
  \BibitemOpen
  \bibfield  {author} {\bibinfo {author} {\bibfnamefont {U.}~\bibnamefont
  {Schneider}}, \bibinfo {author} {\bibfnamefont {L.}~\bibnamefont
  {Hackerm\"uller}}, \bibinfo {author} {\bibfnamefont {S.}~\bibnamefont
  {Will}}, \bibinfo {author} {\bibfnamefont {T.}~\bibnamefont {Best}}, \bibinfo
  {author} {\bibfnamefont {I.}~\bibnamefont {Bloch}}, \bibinfo {author}
  {\bibfnamefont {T.~A.}\ \bibnamefont {Costi}}, \bibinfo {author}
  {\bibfnamefont {R.~W.}\ \bibnamefont {Helmes}}, \bibinfo {author}
  {\bibfnamefont {D.}~\bibnamefont {Rasch}}, \ and\ \bibinfo {author}
  {\bibfnamefont {A.}~\bibnamefont {Rosch}},\ }\href {\doibase
  10.1126/science.1165449} {\bibfield  {journal} {\bibinfo  {journal}
  {Science}\ }\textbf {\bibinfo {volume} {322}},\ \bibinfo {pages} {1520}
  (\bibinfo {year} {2008})}\BibitemShut {NoStop}%
\bibitem [{\citenamefont {Greif}\ \emph {et~al.}(2013)\citenamefont {Greif},
  \citenamefont {Uehlinger}, \citenamefont {Jotzu}, \citenamefont {Tarruell},\
  and\ \citenamefont {Esslinger}}]{Greif2013S}%
  \BibitemOpen
  \bibfield  {author} {\bibinfo {author} {\bibfnamefont {D.}~\bibnamefont
  {Greif}}, \bibinfo {author} {\bibfnamefont {T.}~\bibnamefont {Uehlinger}},
  \bibinfo {author} {\bibfnamefont {G.}~\bibnamefont {Jotzu}}, \bibinfo
  {author} {\bibfnamefont {L.}~\bibnamefont {Tarruell}}, \ and\ \bibinfo
  {author} {\bibfnamefont {T.}~\bibnamefont {Esslinger}},\ }\href {\doibase
  10.1126/science.1236362} {\bibfield  {journal} {\bibinfo  {journal}
  {Science}\ }\textbf {\bibinfo {volume} {340}},\ \bibinfo {pages} {1307}
  (\bibinfo {year} {2013})}\BibitemShut {NoStop}%
\bibitem [{\citenamefont {Hart}\ \emph {et~al.}(2015)\citenamefont {Hart},
  \citenamefont {Duarte}, \citenamefont {Yang}, \citenamefont {Liu},
  \citenamefont {Paiva}, \citenamefont {Khatami}, \citenamefont {Scalettar},
  \citenamefont {Trivedi}, \citenamefont {Huse},\ and\ \citenamefont
  {Hulet}}]{Hart2015Nat}%
  \BibitemOpen
  \bibfield  {author} {\bibinfo {author} {\bibfnamefont {R.~A.}\ \bibnamefont
  {Hart}}, \bibinfo {author} {\bibfnamefont {P.~M.}\ \bibnamefont {Duarte}},
  \bibinfo {author} {\bibfnamefont {T.-L.}\ \bibnamefont {Yang}}, \bibinfo
  {author} {\bibfnamefont {X.}~\bibnamefont {Liu}}, \bibinfo {author}
  {\bibfnamefont {T.}~\bibnamefont {Paiva}}, \bibinfo {author} {\bibfnamefont
  {E.}~\bibnamefont {Khatami}}, \bibinfo {author} {\bibfnamefont {R.~T.}\
  \bibnamefont {Scalettar}}, \bibinfo {author} {\bibfnamefont {N.}~\bibnamefont
  {Trivedi}}, \bibinfo {author} {\bibfnamefont {D.~A.}\ \bibnamefont {Huse}}, \
  and\ \bibinfo {author} {\bibfnamefont {R.~G.}\ \bibnamefont {Hulet}},\ }\href
  {http://dx.doi.org/10.1038/nature14223} {\bibfield  {journal} {\bibinfo
  {journal} {Nature}\ }\textbf {\bibinfo {volume} {519}},\ \bibinfo {pages}
  {211} (\bibinfo {year} {2015})}\BibitemShut {NoStop}%
\bibitem [{\citenamefont {Haller}\ \emph {et~al.}(2015)\citenamefont {Haller},
  \citenamefont {Hudson}, \citenamefont {Kelly}, \citenamefont {Cotta},
  \citenamefont {Peaudecerf}, \citenamefont {Bruce},\ and\ \citenamefont
  {Kuhr}}]{Haller2015}%
  \BibitemOpen
  \bibfield  {author} {\bibinfo {author} {\bibfnamefont {E.}~\bibnamefont
  {Haller}}, \bibinfo {author} {\bibfnamefont {J.}~\bibnamefont {Hudson}},
  \bibinfo {author} {\bibfnamefont {A.}~\bibnamefont {Kelly}}, \bibinfo
  {author} {\bibfnamefont {D.~A.}\ \bibnamefont {Cotta}}, \bibinfo {author}
  {\bibfnamefont {B.}~\bibnamefont {Peaudecerf}}, \bibinfo {author}
  {\bibfnamefont {G.~D.}\ \bibnamefont {Bruce}}, \ and\ \bibinfo {author}
  {\bibfnamefont {S.}~\bibnamefont {Kuhr}},\ }\href@noop {} {\bibfield
  {journal} {\bibinfo  {journal} {Nature Physics}\ }\textbf {\bibinfo {volume}
  {advance online publication}},\  (\bibinfo {year} {2015})}\BibitemShut
  {NoStop}%
\bibitem [{\citenamefont {Cheuk}\ \emph {et~al.}(2015)\citenamefont {Cheuk},
  \citenamefont {Nichols}, \citenamefont {Okan}, \citenamefont {Gersdorf},
  \citenamefont {Ramasesh}, \citenamefont {Bakr}, \citenamefont {Lompe},\ and\
  \citenamefont {Zwierlein}}]{Cheuk2015}%
  \BibitemOpen
  \bibfield  {author} {\bibinfo {author} {\bibfnamefont {L.~W.}\ \bibnamefont
  {Cheuk}}, \bibinfo {author} {\bibfnamefont {M.~A.}\ \bibnamefont {Nichols}},
  \bibinfo {author} {\bibfnamefont {M.}~\bibnamefont {Okan}}, \bibinfo {author}
  {\bibfnamefont {T.}~\bibnamefont {Gersdorf}}, \bibinfo {author}
  {\bibfnamefont {V.~V.}\ \bibnamefont {Ramasesh}}, \bibinfo {author}
  {\bibfnamefont {W.~S.}\ \bibnamefont {Bakr}}, \bibinfo {author}
  {\bibfnamefont {T.}~\bibnamefont {Lompe}}, \ and\ \bibinfo {author}
  {\bibfnamefont {M.~W.}\ \bibnamefont {Zwierlein}},\ }\href {\doibase
  10.1103/PhysRevLett.114.193001} {\bibfield  {journal} {\bibinfo  {journal}
  {Phys. Rev. Lett.}\ }\textbf {\bibinfo {volume} {114}},\ \bibinfo {pages}
  {193001} (\bibinfo {year} {2015})}\BibitemShut {NoStop}%
\bibitem [{\citenamefont {Parsons}\ \emph {et~al.}(2015)\citenamefont
  {Parsons}, \citenamefont {Huber}, \citenamefont {Mazurenko}, \citenamefont
  {Chiu}, \citenamefont {Setiawan}, \citenamefont {Wooley-Brown}, \citenamefont
  {Blatt},\ and\ \citenamefont {Greiner}}]{Parsons2015}%
  \BibitemOpen
  \bibfield  {author} {\bibinfo {author} {\bibfnamefont {M.~F.}\ \bibnamefont
  {Parsons}}, \bibinfo {author} {\bibfnamefont {F.}~\bibnamefont {Huber}},
  \bibinfo {author} {\bibfnamefont {A.}~\bibnamefont {Mazurenko}}, \bibinfo
  {author} {\bibfnamefont {C.~S.}\ \bibnamefont {Chiu}}, \bibinfo {author}
  {\bibfnamefont {W.}~\bibnamefont {Setiawan}}, \bibinfo {author}
  {\bibfnamefont {K.}~\bibnamefont {Wooley-Brown}}, \bibinfo {author}
  {\bibfnamefont {S.}~\bibnamefont {Blatt}}, \ and\ \bibinfo {author}
  {\bibfnamefont {M.}~\bibnamefont {Greiner}},\ }\href {\doibase
  10.1103/PhysRevLett.114.213002} {\bibfield  {journal} {\bibinfo  {journal}
  {Phys. Rev. Lett.}\ }\textbf {\bibinfo {volume} {114}},\ \bibinfo {pages}
  {213002} (\bibinfo {year} {2015})}\BibitemShut {NoStop}%
\bibitem [{\citenamefont {Esslinger}(2010)}]{Ess2010ARCMP}%
  \BibitemOpen
  \bibfield  {author} {\bibinfo {author} {\bibfnamefont {T.}~\bibnamefont
  {Esslinger}},\ }\href {\doibase 10.1146/annurev-conmatphys-070909-104059}
  {\bibfield  {journal} {\bibinfo  {journal} {Annu. Rev. Condens. Matter
  Phys.}\ }\textbf {\bibinfo {volume} {1}},\ \bibinfo {pages} {129} (\bibinfo
  {year} {2010})}\BibitemShut {NoStop}%
\bibitem [{\citenamefont {Bloch}\ \emph {et~al.}(2008)\citenamefont {Bloch},
  \citenamefont {Dalibard},\ and\ \citenamefont {Zwerger}}]{Blo2008RMP}%
  \BibitemOpen
  \bibfield  {author} {\bibinfo {author} {\bibfnamefont {I.}~\bibnamefont
  {Bloch}}, \bibinfo {author} {\bibfnamefont {J.}~\bibnamefont {Dalibard}}, \
  and\ \bibinfo {author} {\bibfnamefont {W.}~\bibnamefont {Zwerger}},\ }\href
  {\doibase 10.1103/RevModPhys.80.885} {\bibfield  {journal} {\bibinfo
  {journal} {Rev. Mod. Phys.}\ }\textbf {\bibinfo {volume} {80}},\ \bibinfo
  {pages} {885} (\bibinfo {year} {2008})}\BibitemShut {NoStop}%
\bibitem [{\citenamefont {Imri\ifmmode~\check{s}\else \v{s}\fi{}ka}\ \emph
  {et~al.}(2014)\citenamefont {Imri\ifmmode~\check{s}\else \v{s}\fi{}ka},
  \citenamefont {Iazzi}, \citenamefont {Wang}, \citenamefont {Gull},
  \citenamefont {Greif}, \citenamefont {Uehlinger}, \citenamefont {Jotzu},
  \citenamefont {Tarruell}, \citenamefont {Esslinger},\ and\ \citenamefont
  {Troyer}}]{Imriska2014}%
  \BibitemOpen
  \bibfield  {author} {\bibinfo {author} {\bibfnamefont {J.}~\bibnamefont
  {Imri\ifmmode~\check{s}\else \v{s}\fi{}ka}}, \bibinfo {author} {\bibfnamefont
  {M.}~\bibnamefont {Iazzi}}, \bibinfo {author} {\bibfnamefont
  {L.}~\bibnamefont {Wang}}, \bibinfo {author} {\bibfnamefont {E.}~\bibnamefont
  {Gull}}, \bibinfo {author} {\bibfnamefont {D.}~\bibnamefont {Greif}},
  \bibinfo {author} {\bibfnamefont {T.}~\bibnamefont {Uehlinger}}, \bibinfo
  {author} {\bibfnamefont {G.}~\bibnamefont {Jotzu}}, \bibinfo {author}
  {\bibfnamefont {L.}~\bibnamefont {Tarruell}}, \bibinfo {author}
  {\bibfnamefont {T.}~\bibnamefont {Esslinger}}, \ and\ \bibinfo {author}
  {\bibfnamefont {M.}~\bibnamefont {Troyer}},\ }\href {\doibase
  10.1103/PhysRevLett.112.115301} {\bibfield  {journal} {\bibinfo  {journal}
  {Phys. Rev. Lett.}\ }\textbf {\bibinfo {volume} {112}},\ \bibinfo {pages}
  {115301} (\bibinfo {year} {2014})}\BibitemShut {NoStop}%
\bibitem [{\citenamefont {Georges}\ \emph {et~al.}(1996)\citenamefont
  {Georges}, \citenamefont {Kotliar}, \citenamefont {Krauth},\ and\
  \citenamefont {Rozenberg}}]{Geo1996RMP}%
  \BibitemOpen
  \bibfield  {author} {\bibinfo {author} {\bibfnamefont {A.}~\bibnamefont
  {Georges}}, \bibinfo {author} {\bibfnamefont {G.}~\bibnamefont {Kotliar}},
  \bibinfo {author} {\bibfnamefont {W.}~\bibnamefont {Krauth}}, \ and\ \bibinfo
  {author} {\bibfnamefont {M.~J.}\ \bibnamefont {Rozenberg}},\ }\href {\doibase
  10.1103/RevModPhys.68.13} {\bibfield  {journal} {\bibinfo  {journal} {Rev.
  Mod. Phys.}\ }\textbf {\bibinfo {volume} {68}},\ \bibinfo {pages} {13}
  (\bibinfo {year} {1996})}\BibitemShut {NoStop}%
\bibitem [{\citenamefont {Anderson}(1961)}]{Anderson1961}%
  \BibitemOpen
  \bibfield  {author} {\bibinfo {author} {\bibfnamefont {P.~W.}\ \bibnamefont
  {Anderson}},\ }\href {\doibase 10.1103/PhysRev.124.41} {\bibfield  {journal}
  {\bibinfo  {journal} {Phys. Rev.}\ }\textbf {\bibinfo {volume} {124}},\
  \bibinfo {pages} {41} (\bibinfo {year} {1961})}\BibitemShut {NoStop}%
\bibitem [{\citenamefont {Caffarel}\ and\ \citenamefont
  {Krauth}(1994)}]{Caffarel1994PRL}%
  \BibitemOpen
  \bibfield  {author} {\bibinfo {author} {\bibfnamefont {M.}~\bibnamefont
  {Caffarel}}\ and\ \bibinfo {author} {\bibfnamefont {W.}~\bibnamefont
  {Krauth}},\ }\href {\doibase 10.1103/PhysRevLett.72.1545} {\bibfield
  {journal} {\bibinfo  {journal} {Phys. Rev. Lett.}\ }\textbf {\bibinfo
  {volume} {72}},\ \bibinfo {pages} {1545} (\bibinfo {year}
  {1994})}\BibitemShut {NoStop}%
\bibitem [{\citenamefont {Sotnikov}\ \emph {et~al.}(2013)\citenamefont
  {Sotnikov}, \citenamefont {Snoek},\ and\ \citenamefont
  {Hofstetter}}]{Sotnikov2013}%
  \BibitemOpen
  \bibfield  {author} {\bibinfo {author} {\bibfnamefont {A.}~\bibnamefont
  {Sotnikov}}, \bibinfo {author} {\bibfnamefont {M.}~\bibnamefont {Snoek}}, \
  and\ \bibinfo {author} {\bibfnamefont {W.}~\bibnamefont {Hofstetter}},\
  }\href {\doibase 10.1103/PhysRevA.87.053602} {\bibfield  {journal} {\bibinfo
  {journal} {Phys. Rev. A}\ }\textbf {\bibinfo {volume} {87}},\ \bibinfo
  {pages} {053602} (\bibinfo {year} {2013})}\BibitemShut {NoStop}%
\bibitem [{\citenamefont {Mermin}\ and\ \citenamefont
  {Wagner}(1966)}]{MerminWagner1966}%
  \BibitemOpen
  \bibfield  {author} {\bibinfo {author} {\bibfnamefont {N.~D.}\ \bibnamefont
  {Mermin}}\ and\ \bibinfo {author} {\bibfnamefont {H.}~\bibnamefont
  {Wagner}},\ }\href {\doibase 10.1103/PhysRevLett.17.1133} {\bibfield
  {journal} {\bibinfo  {journal} {Phys. Rev. Lett.}\ }\textbf {\bibinfo
  {volume} {17}},\ \bibinfo {pages} {1133} (\bibinfo {year}
  {1966})}\BibitemShut {NoStop}%
\bibitem [{\citenamefont {Gorelik}\ and\ \citenamefont
  {Bl\"umer}(2011)}]{Gorelik2011JLTP}%
  \BibitemOpen
  \bibfield  {author} {\bibinfo {author} {\bibfnamefont {E.}~\bibnamefont
  {Gorelik}}\ and\ \bibinfo {author} {\bibfnamefont {N.}~\bibnamefont
  {Bl\"umer}},\ }\href {\doibase 10.1007/s10909-011-0396-3} {\bibfield
  {journal} {\bibinfo  {journal} {Journal of Low Temperature Physics}\ }\textbf
  {\bibinfo {volume} {165}},\ \bibinfo {pages} {195} (\bibinfo {year}
  {2011})}\BibitemShut {NoStop}%
\bibitem [{\citenamefont {van Dongen}(1991)}]{vanDongen}%
  \BibitemOpen
  \bibfield  {author} {\bibinfo {author} {\bibfnamefont {P.~G.~J.}\
  \bibnamefont {van Dongen}},\ }\href@noop {} {\bibfield  {journal} {\bibinfo
  {journal} {Phys. Rev. Lett.}\ }\textbf {\bibinfo {volume} {67}},\ \bibinfo
  {pages} {757 } (\bibinfo {year} {1991})}\BibitemShut {NoStop}%
\bibitem [{\citenamefont {Scarola}\ \emph {et~al.}(2009)\citenamefont
  {Scarola}, \citenamefont {Pollet}, \citenamefont {Oitmaa},\ and\
  \citenamefont {Troyer}}]{Scarola2009}%
  \BibitemOpen
  \bibfield  {author} {\bibinfo {author} {\bibfnamefont {V.~W.}\ \bibnamefont
  {Scarola}}, \bibinfo {author} {\bibfnamefont {L.}~\bibnamefont {Pollet}},
  \bibinfo {author} {\bibfnamefont {J.}~\bibnamefont {Oitmaa}}, \ and\ \bibinfo
  {author} {\bibfnamefont {M.}~\bibnamefont {Troyer}},\ }\href {\doibase
  10.1103/PhysRevLett.102.135302} {\bibfield  {journal} {\bibinfo  {journal}
  {Phys. Rev. Lett.}\ }\textbf {\bibinfo {volume} {102}},\ \bibinfo {pages}
  {135302} (\bibinfo {year} {2009})}\BibitemShut {NoStop}%
\bibitem [{\citenamefont {Gorelik}\ \emph {et~al.}(2010)\citenamefont
  {Gorelik}, \citenamefont {Titvinidze}, \citenamefont {Hofstetter},
  \citenamefont {Snoek},\ and\ \citenamefont {Bl\"umer}}]{Gor2010PRL}%
  \BibitemOpen
  \bibfield  {author} {\bibinfo {author} {\bibfnamefont {E.~V.}\ \bibnamefont
  {Gorelik}}, \bibinfo {author} {\bibfnamefont {I.}~\bibnamefont {Titvinidze}},
  \bibinfo {author} {\bibfnamefont {W.}~\bibnamefont {Hofstetter}}, \bibinfo
  {author} {\bibfnamefont {M.}~\bibnamefont {Snoek}}, \ and\ \bibinfo {author}
  {\bibfnamefont {N.}~\bibnamefont {Bl\"umer}},\ }\href {\doibase
  10.1103/PhysRevLett.105.065301} {\bibfield  {journal} {\bibinfo  {journal}
  {Phys. Rev. Lett.}\ }\textbf {\bibinfo {volume} {105}},\ \bibinfo {pages}
  {065301} (\bibinfo {year} {2010})}\BibitemShut {NoStop}%
\bibitem [{\citenamefont {Khatami}\ and\ \citenamefont
  {Rigol}(2011)}]{Khatami2011}%
  \BibitemOpen
  \bibfield  {author} {\bibinfo {author} {\bibfnamefont {E.}~\bibnamefont
  {Khatami}}\ and\ \bibinfo {author} {\bibfnamefont {M.}~\bibnamefont
  {Rigol}},\ }\href {\doibase 10.1103/PhysRevA.84.053611} {\bibfield  {journal}
  {\bibinfo  {journal} {Phys. Rev. A}\ }\textbf {\bibinfo {volume} {84}},\
  \bibinfo {pages} {053611} (\bibinfo {year} {2011})}\BibitemShut {NoStop}%
\bibitem [{\citenamefont {Fuchs}\ \emph {et~al.}(2011)\citenamefont {Fuchs},
  \citenamefont {Gull}, \citenamefont {Pollet}, \citenamefont {Burovski},
  \citenamefont {Kozik}, \citenamefont {Pruschke},\ and\ \citenamefont
  {Troyer}}]{Fuchs2011}%
  \BibitemOpen
  \bibfield  {author} {\bibinfo {author} {\bibfnamefont {S.}~\bibnamefont
  {Fuchs}}, \bibinfo {author} {\bibfnamefont {E.}~\bibnamefont {Gull}},
  \bibinfo {author} {\bibfnamefont {L.}~\bibnamefont {Pollet}}, \bibinfo
  {author} {\bibfnamefont {E.}~\bibnamefont {Burovski}}, \bibinfo {author}
  {\bibfnamefont {E.}~\bibnamefont {Kozik}}, \bibinfo {author} {\bibfnamefont
  {T.}~\bibnamefont {Pruschke}}, \ and\ \bibinfo {author} {\bibfnamefont
  {M.}~\bibnamefont {Troyer}},\ }\href {\doibase
  10.1103/PhysRevLett.106.030401} {\bibfield  {journal} {\bibinfo  {journal}
  {Phys. Rev. Lett.}\ }\textbf {\bibinfo {volume} {106}},\ \bibinfo {pages}
  {030401} (\bibinfo {year} {2011})}\BibitemShut {NoStop}%
\bibitem [{\citenamefont {Schneider}(2010)}]{SchneiderThesis}%
  \BibitemOpen
  \bibfield  {author} {\bibinfo {author} {\bibfnamefont {U.}~\bibnamefont
  {Schneider}},\ }\emph {\bibinfo {title} {Interacting Fermionic Atoms in
  Optical Lattices - A Quantum Simulator for Condensed Matter Physics}},\
  \href@noop {} {Ph.D. thesis},\ \bibinfo  {school} {Johannes
  Gutenberg-Universit\"at in Mainz} (\bibinfo {year} {2010})\BibitemShut
  {NoStop}%
\bibitem [{\citenamefont {Werner}\ \emph {et~al.}(2005)\citenamefont {Werner},
  \citenamefont {Parcollet}, \citenamefont {Georges},\ and\ \citenamefont
  {Hassan}}]{Wer2005PRL}%
  \BibitemOpen
  \bibfield  {author} {\bibinfo {author} {\bibfnamefont {F.}~\bibnamefont
  {Werner}}, \bibinfo {author} {\bibfnamefont {O.}~\bibnamefont {Parcollet}},
  \bibinfo {author} {\bibfnamefont {A.}~\bibnamefont {Georges}}, \ and\
  \bibinfo {author} {\bibfnamefont {S.~R.}\ \bibnamefont {Hassan}},\ }\href
  {\doibase 10.1103/PhysRevLett.95.056401} {\bibfield  {journal} {\bibinfo
  {journal} {Phys. Rev. Lett.}\ }\textbf {\bibinfo {volume} {95}},\ \bibinfo
  {pages} {056401} (\bibinfo {year} {2005})}\BibitemShut {NoStop}%
\bibitem [{\citenamefont {Kozik}\ \emph {et~al.}(2013)\citenamefont {Kozik},
  \citenamefont {Burovski}, \citenamefont {Scarola},\ and\ \citenamefont
  {Troyer}}]{Kozik2013}%
  \BibitemOpen
  \bibfield  {author} {\bibinfo {author} {\bibfnamefont {E.}~\bibnamefont
  {Kozik}}, \bibinfo {author} {\bibfnamefont {E.}~\bibnamefont {Burovski}},
  \bibinfo {author} {\bibfnamefont {V.~W.}\ \bibnamefont {Scarola}}, \ and\
  \bibinfo {author} {\bibfnamefont {M.}~\bibnamefont {Troyer}},\ }\href
  {\doibase 10.1103/PhysRevB.87.205102} {\bibfield  {journal} {\bibinfo
  {journal} {Phys. Rev. B}\ }\textbf {\bibinfo {volume} {87}},\ \bibinfo
  {pages} {205102} (\bibinfo {year} {2013})}\BibitemShut {NoStop}%
\bibitem [{\citenamefont {Bakr}\ \emph {et~al.}(2009)\citenamefont {Bakr},
  \citenamefont {Gillen}, \citenamefont {Peng}, \citenamefont {F\"olling},\
  and\ \citenamefont {Greiner}}]{Bakr2009Nat}%
  \BibitemOpen
  \bibfield  {author} {\bibinfo {author} {\bibfnamefont {W.~S.}\ \bibnamefont
  {Bakr}}, \bibinfo {author} {\bibfnamefont {J.~I.}\ \bibnamefont {Gillen}},
  \bibinfo {author} {\bibfnamefont {A.}~\bibnamefont {Peng}}, \bibinfo {author}
  {\bibfnamefont {S.}~\bibnamefont {F\"olling}}, \ and\ \bibinfo {author}
  {\bibfnamefont {M.}~\bibnamefont {Greiner}},\ }\href
  {http://dx.doi.org/10.1038/nature08482} {\bibfield  {journal} {\bibinfo
  {journal} {Nature}\ }\textbf {\bibinfo {volume} {462}},\ \bibinfo {pages}
  {74} (\bibinfo {year} {2009})}\BibitemShut {NoStop}%
\bibitem [{\citenamefont {Sherson}\ \emph {et~al.}(2010)\citenamefont
  {Sherson}, \citenamefont {Weitenberg}, \citenamefont {Endres}, \citenamefont
  {Cheneau}, \citenamefont {Bloch},\ and\ \citenamefont {Kuhr}}]{Sherson2010}%
  \BibitemOpen
  \bibfield  {author} {\bibinfo {author} {\bibfnamefont {J.~F.}\ \bibnamefont
  {Sherson}}, \bibinfo {author} {\bibfnamefont {C.}~\bibnamefont {Weitenberg}},
  \bibinfo {author} {\bibfnamefont {M.}~\bibnamefont {Endres}}, \bibinfo
  {author} {\bibfnamefont {M.}~\bibnamefont {Cheneau}}, \bibinfo {author}
  {\bibfnamefont {I.}~\bibnamefont {Bloch}}, \ and\ \bibinfo {author}
  {\bibfnamefont {S.}~\bibnamefont {Kuhr}},\ }\href@noop {} {\bibfield
  {journal} {\bibinfo  {journal} {Nature}\ }\textbf {\bibinfo {volume} {467}},\
  \bibinfo {pages} {68} (\bibinfo {year} {2010})}\BibitemShut {NoStop}%
\end{thebibliography}%
\appendix
\section{Non-interacting density of states for anisotropic systems}\label{app.A}

The general expression for the density of states (DOS) in $n$ dimensions is given by
\begin{eqnarray}\label{Eq:Dos}
D_{n}(\varepsilon) = \frac{1}{(2\pi)^{n}} \int \limits_{BZ} d^{n}k ~ \delta(\varepsilon-\varepsilon(\bf{k}))
\end{eqnarray}
with integration over the Brillouin zone.
The energy is a function of the wave vector $\bf{k}$. For cubic lattice geometry its general form is given by
\begin{displaymath}
\varepsilon(\textbf{k}) = - 2 \sum \limits_{\alpha} t_{\alpha} \cos(k_{\alpha}a_{\alpha}),
\end{displaymath}
where $t_{\alpha}$ denotes the hopping parameter and $a_{\alpha}$ the lattice spacing in $\alpha$-direction, and $k_{\alpha}$ are the corresponding components of the wave vector. We consider lattice geometries with equal lattice spacings in all directions and set $a_{\alpha}=1$ for simplicity.

Our aim is to calculate the non-interacting density of states for anisotropic hopping, i.e. different values of $t_\alpha$ for $\alpha = x, y, z$, in particular, for the case of isotropic hopping in the $xy$-plane and varying the hopping parameter in $z$-direction.
In the following, the closed analytical expression for the density of states is derived in terms of complete elliptic integrals of the first kind.


\subsection{DOS in One Dimension}
In one dimension the energy and the density of states are given by
\begin{eqnarray*}
\varepsilon(k_{x}) = -2 t_{x} \cos(k_{x}) \equiv -\varepsilon_{0} \cos(k_{x}),\\
D_{1}(\varepsilon) = \frac{1}{2\pi} \int \limits_{-\pi}^{\pi} dk_{x}\,\delta(\varepsilon-\varepsilon(k_{x})).
\end{eqnarray*}
The integral is evaluated using the following property of the Delta function:
\begin{eqnarray}\label{Eq:delta}
\int \limits_{-\infty}^{\infty} dx f(x) \delta(g(x)) = \sum \limits_{i} \frac{f(x_{i})}{|g'(x_{i})|} \\Ê\nonumber ~Ê\text{with} ~Êg(x_{i}) = 0.
\end{eqnarray}

Due to the periodicity of the system the integration limits in the integral for $D_1$ can be extended to infinity and the equation for the DOS reads
\begin{align*}
D_{1}(\varepsilon) &= \frac{1}{2\pi} \sum \limits_{i} \frac{1}{|-\varepsilon_{0} \sin(\arccos(-\varepsilon/\varepsilon_{0}))|}\\
&= \frac{1}{\pi} \frac{1}{|-\varepsilon_{0} \sin(\arccos(-\varepsilon/\varepsilon_{0}))|}.
\end{align*}

The sum just gives a factor of 2, since the two roots of the function $g(k_{x})=\varepsilon+\varepsilon_{0} \cos(k_{x})$ are given by $\pm \arccos(-\varepsilon/\varepsilon_{0})$, in accordance with the symmetry of the cosine function. In the last step we use the relation
\begin{eqnarray}\label{Eq:sinarccos}
\sin(\arccos(x)) = \cos(\arcsin(x)) = \sqrt{1-x^{2}},
\end{eqnarray}
and the final result reads:
\begin{eqnarray*}
D_{1}(\varepsilon) = \frac{1}{\pi\sqrt{\varepsilon_{0}^{2}-\varepsilon^{2}}}.
\end{eqnarray*}


\subsection{DOS in Two Dimensions}
In two dimensions, we consider a square lattice with anisotropic hopping $t_{x} \neq t_{y}$. The energy is given by
\begin{eqnarray*}
\varepsilon(k_{x}, k_{y}) = -2t_{x} \cos(k_{x}) - 2t_{y} \cos(k_{y})
\end{eqnarray*}
and the density of states reads
\begin{align*}
D_{2}(\varepsilon) = \frac{1}{(2\pi)^{2}} \int \limits_{-\pi}^{\pi} dk_{x} \int \limits_{-\pi}^{\pi} dk_{y}~ \delta(\varepsilon - \varepsilon(k_{x}, k_{y}))\\ 
= \frac{1}{(2\pi)^{2}} \int \limits_{-\pi}^{\pi} dk_{x} \int \limits_{-\pi}^{\pi} dk_{y}~ \delta(\varepsilon + 2t_{x} \cos(k_{x}) + 2t_{y} \cos(k_{y}))\\
= \frac{1}{\pi^{2}2t_{y}} \int \limits_{0}^{\pi} dk_{x} \int \limits_{0}^{\pi} dk_{y}~ \delta(\tilde\varepsilon + \Delta \cos(k_{x}) + \cos(k_{y})),
\end{align*}
where in the last step we introduce dimensionless parameters $\tilde\varepsilon \equiv \varepsilon/2t_{y}$ and $\Delta \equiv t_{x}/t_{y}$ and change the integration limits according to the symmetry of the integrand. In the next step the substitution $\alpha = \cos(k_{\alpha})$ is performed for $\alpha = x, y$ and the DOS becomes:
\begin{eqnarray*}
D_{2}(\varepsilon) = \frac{1}{\pi^{2}2t_{y}} \int \limits_{-1}^{1} dx \int \limits_{-1}^{1} dy~ \frac{\delta(\tilde\varepsilon + x \Delta + y)}{\sqrt{(1-x^2)(1-y^2)}}.
\end{eqnarray*}
The evaluation of the $\delta$-function for $y$ according to Eq.~(\ref{Eq:delta}) gives
\begin{eqnarray*}
D_{2}(\varepsilon) = \frac{1}{\pi^{2}2t_{y}} \int dx ~ \frac{1}{\sqrt{(1-x^2)[1-(\tilde\varepsilon + x\Delta)^2]}}.
\end{eqnarray*}

The integration limits of $x$ are changed by this transformation and are now determined by the square root in the denominator. Using the relation (\ref{Eq:sinarccos}) we obtain $\sqrt{(1-x^2)[1-(\tilde\varepsilon + x\Delta)^2]}=\sin(\arccos(x))\sin(\arccos(\tilde\varepsilon + x\Delta))$.
The argument of $\arccos$ is restricted to values in the interval [-1, 1], thus the two conditions
\begin{displaymath} -1\leq x \leq 1 \quad ~\text{ and } \quad -1\leq (\tilde\varepsilon + x\Delta) \leq 1\end{displaymath}
have to be satisfied. This leads to the following limits:
\begin{eqnarray*}
x_{min} = \max\bigg(-1, \frac{-1-\tilde\varepsilon}{\Delta}\bigg), \quad x_{max} = \min\bigg(1, \frac{1-\tilde\varepsilon}{\Delta}\bigg)
\end{eqnarray*}
with $-1-\Delta \leq \tilde\varepsilon \leq 1+\Delta$.
Evaluation of the dependence on $\tilde\varepsilon$ leads to four distinct regions:
\begin{center}
\renewcommand{\arraystretch}{1.8}
\begin{tabular}{l|rl|l}
$\tilde\varepsilon > 0$ 	& I: & $1-\Delta < \tilde\varepsilon < 1+\Delta$ 	& $-1 \leq x \leq \frac{1-\tilde\varepsilon}{\Delta}$ \\
				 	& II: & $0 < \tilde\varepsilon < 1- \Delta$ 		& $-1 \leq x \leq 1$\\ \hline
$\tilde\varepsilon < 0$Ê	& III: & $\Delta-1 < \tilde\varepsilon < 0$ 		& $-1 \leq x \leq 1$\\
					& IV: & $0 < \tilde\varepsilon < \Delta-1$		& $\frac{-1-\tilde\varepsilon}{\Delta} \leq x \leq 1$
\end{tabular}
\renewcommand{\arraystretch}{1}
\end{center}
But due to the symmetry of the integrand we find that $D^{I}(\tilde\varepsilon > 0; \Delta) = D^{IV}(\tilde\varepsilon < 0; \Delta)$ and $D^{II}(\tilde\varepsilon > 0; \Delta) = D^{III}(\tilde\varepsilon < 0; \Delta)$. This result will be considered after the following transformation.

To evaluate the integral we factorize the polynome under the square root in $x$, bringing it to the form $P(x)=\prod_{i} (x-x_{i})$, with $x_{i}$ being the roots of the polynome,
\begin{eqnarray*}
x_{1} = 1, ~Ê~Ê x_{2} = -1, ~ ~Êx_{3} = \frac{1-\tilde\varepsilon}{\Delta}, ~Ê~Êx_{4} = \frac{-1-\tilde\varepsilon}{\Delta},
\end{eqnarray*}
so that the DOS reads:
\begin{eqnarray*}
D_{2}(\varepsilon) = \frac{1}{\pi^{2}2t_{y}} \int dx ~ \frac{1}{\sqrt{P(x)}}.
\end{eqnarray*}

In the next step, the following substitution is applied to the factorized polynome:
\begin{eqnarray}\label{Eq:substitution}
x(\phi) = \frac{\gamma(\beta-\delta) - \delta(\beta-\gamma) \sin^{2}(\phi)}{(\beta-\delta) - (\beta-\gamma) \sin^{2}(\phi)}
\end{eqnarray}
with $\delta < \gamma < \beta < \alpha$ being the real roots of the polynome. This way, the integral is transformed into an elliptic integral of the first kind:
\begin{eqnarray*}
\int \frac{dx}{\sqrt{P(x)}} ~Ê~Ê\longrightarrow ~ ~Ê\frac{2}{\sqrt{(\alpha-\gamma)(\beta-\delta)}} \int \frac{d\phi}{\sqrt{1-m\sin^{2}(\phi)}}
\end{eqnarray*}
with the parameter $m$ given by
\begin{eqnarray*}
m = \frac{(\beta-\gamma)(\alpha-\delta)}{(\alpha-\gamma)(\beta-\delta)}, \quad \gamma < x < \beta.
\end{eqnarray*}
The integration limits transform according to the formula 
\begin{eqnarray*}
\phi(x) = \arcsin\left(\sqrt{\frac{(x-\gamma)(\beta-\delta)}{(x-\delta)(\beta-\gamma)}}\right).
\end{eqnarray*}

This transformation is performed separately in the two regions stated above, i.e. for $D^{I}=D^{IV}$ and $D^{II} = D^{III}$, since the grading of the roots $x_{i}$ is dependent on $\tilde\varepsilon$. For the regions I and IV the grading is given by $x_{1}>x_{3}>x_{2}>x_{4}$ and the DOS transforms into
\begin{eqnarray*}
D_{2}(\varepsilon) = \frac{\sqrt{\Delta}}{\pi^{2}2t_{x}} K(m_{a}) , ~Ê~Ê~Êm_{a} = \frac{(1+\Delta)^2-\tilde\varepsilon^2}{4\Delta}
\end{eqnarray*}
and for the regions II and III we obtain $x_{3}>x_{1}>x_{2}>x_{4}$ with the DOS given by
\begin{eqnarray*}
D_{2}(\varepsilon) = \frac{1}{\pi^{2}2t_{x}} \frac{2\Delta \cdot K(m_{b})}{\sqrt{(1+\Delta)^2+\tilde\varepsilon}} , ~Ê~Ê~Êm_{b} = \frac{4\Delta}{(1+\Delta)^2-\tilde\varepsilon^2},
\end{eqnarray*}
where $K(m)$ denotes a complete elliptic integral of the first kind with the parameter $m$, given in the Legendre form by
\begin{eqnarray}\label{Eq:Elli}
K(m) = \int \limits_{0}^{\pi/2} \frac{d\phi}{\sqrt{1-m\sin^{2}(\phi)}}.
\end{eqnarray}


\subsection{DOS in Three Dimensions}
In three dimensions we consider a simple cubic lattice with equal hopping parameters in the $xy$-plane, $t_{z}~\neq~t_{x}~=~t_{y}~\equiv~t$. The kinetic energy is given by
\begin{eqnarray*}
\varepsilon(k_{x}, k_{y}, k_{z}) = -2t[\cos(k_{x}) + \cos(k_{y})] - 2t_{z} \cos(k_{y})
\end{eqnarray*}
and the density of states is
\begin{align*}
D_{3}(\varepsilon) &= \frac{1}{(2\pi)^{3}} \int \limits_{-\pi}^{\pi} dk_{x} \int \limits_{-\pi}^{\pi} dk_{y} \int \limits_{-\pi}^{\pi} dk_{z} ~Ê\delta(\varepsilon-\varepsilon(k_{x}, k_{y}, k_{z}))\\
&= \frac{1}{\pi^{3}2t} \int \limits_{0}^{\pi} dk_{x} \int \limits_{0}^{\pi} dk_{y} \int \limits_{0}^{\pi} dk_{z} \\
& \qquad\qquad\times \delta\big(\tilde\varepsilon+\cos(k_{x})+\cos(k_{y})+\Delta\cos(k_{z})\big),
\end{align*}
where we introduce the dimensionless parameters $\tilde\varepsilon \equiv \varepsilon/2t$ and $\Delta \equiv t_{z}/t$ and reduce the integration range using the symmetry of the integral. Proceeding in an analogous way as in 2d, we first substitute the cosine terms by $\alpha = \cos(k_{\alpha})$ for $\alpha = x, y, z$ to obtain
\begin{eqnarray*}
D_{3}(\varepsilon) = \frac{1}{\pi^{3}2t} \int \limits_{-1}^{1} dx \int \limits_{-1}^{1} dy \int \limits_{-1}^{1} dz ~Ê\frac{\delta(\tilde\varepsilon+x+y+z \Delta)}{\sqrt{(1-x^2)(1-y^2)(1-z^2)}}
\end{eqnarray*}
and evaluate the $\delta$-function for $y$ according to Eq. (\ref{Eq:delta}), taking $x$ and $z$ as constants, which leads to
\begin{eqnarray*}
D_{3}(\varepsilon) =\\
\frac{1}{\pi^{3}2t} \int dx \int dz ~Ê\frac{1}{\sqrt{(1-x^2)(1-z^2)[1-(\tilde\varepsilon+x+z\Delta)^{2}]}}.
\end{eqnarray*}

Expressing the denominator in terms of the relation~(\ref{Eq:sinarccos}) and considering the domain of $\arccos(x)$ as shown in the previous section, we find the following integration limits:
\begin{align*}
-1 \leq &z \leq 1,\quad x_{min}\leq x\leq x_{max}~, \\
x_{min} &= \max[-1, -1-(\tilde\varepsilon+z\Delta)], \\
x_{max} &= \min[1, 1-(\tilde\varepsilon+z\Delta)].
\end{align*}
The dependence on $(\tilde\varepsilon+z\Delta)$ leads to two distinct regions that need to be considered separately in the following discussion:
\begin{center}
\renewcommand{\arraystretch}{1.8}
\begin{tabular}{c|cc}
$(\tilde\varepsilon+z\Delta) > 0$ 	& I: 	& $-1 < x < 1-(\tilde\varepsilon+z\Delta)$ \\ \hline
$(\tilde\varepsilon+z\Delta) < 0$Ê	& II: 	& $-1-(\tilde\varepsilon+z\Delta) < x < 1$
\end{tabular}
\renewcommand{\arraystretch}{1}
\end{center}

Next, the polynome under the square root is factorized in $x$. The roots of the polynome function are
\begin{eqnarray*}
x_{1} = 1, ~Ê~Ê x_{2} = -1, ~ ~Êx_{3} =  -1-\tilde\varepsilon - z\Delta, ~Ê~Êx_{4} = 1-\tilde\varepsilon - z\Delta ,
\end{eqnarray*}
and the DOS becomes
\begin{eqnarray*}
D_{3}(\varepsilon) = \frac{1}{\pi^{3}2t} \int dx \int dz ~Ê\frac{1}{\sqrt{(1-z^2)\prod \limits_{i=1}^{4} (x-x_{i})}}.
\end{eqnarray*}

Now we perform a substitution given in Eq. (\ref{Eq:substitution}) to transform the DOS into
\begin{eqnarray*}
D_{3}(\varepsilon) = \frac{1}{\pi^{3}2t} \intÊ\frac{dz}{\sqrt{(1-z^2)}} \frac{2}{\sqrt{(\alpha-\gamma)(\beta-\delta)}} \times \\ \int \frac{d\phi}{\sqrt{1-m\sin^{2}(\phi)}}.
\end{eqnarray*}
Here, the two regions shown above need to be considered separately. In the region I the grading of the roots reads $x_{1}>x_{4}>x_{2}>x_{3}$ and in region II we find $x_{4}>x_{1}>x_{3}>x_{2}$, but both lead to the same transformation, thus the DOS is given by
\begin{eqnarray*}
D_{3}(\varepsilon) = \frac{1}{\pi^{3}2t} \int \limits_{-1}^{1} \frac{dz}{\sqrt{1-z^2}} K(m) , ~Ê~Ê~ m = 1-\left(\frac{\tilde\varepsilon+z\Delta}{2}\right)^2
\end{eqnarray*}
with the compete elliptic integral $K(m)$ defined in Eq.~(\ref{Eq:Elli}).

\section{The AGM method}\label{app.B}
Complete elliptic integrals of the first kind can be calculated exactly by the Gaussian AGM method.

It was found by Gauss that the sequences for geometric and arithmetic mean have the same limit -- the arithmetic-geometric mean. Given two starting values $a_{0}$ and $b_{0}$, the sequences read:
\begin{displaymath}
a_{1} = \frac{1}{2} (a_{0}+b_{0}) \,, \, ... \, , \, \, a_{N+1} = \frac{1}{2}(a_{N} + b_{N})
\end{displaymath}
and
\begin{displaymath}
b_{1} = \sqrt{a_{0}b_{0}} \,, \, ... \, , \, \, b_{N+1} = \sqrt{a_{N} b_{N}}.
\end{displaymath}
In the limit $N \to \infty$ both sequences converge to the same value:
\begin{displaymath}
\lim_{N \to \infty} a_{N} = \lim_{N \to \infty} b_{N} = agm(a, b).
\end{displaymath}

For elliptic integrals of the first kind, Gauss derived the simple relation
\begin{displaymath}
I(a, b) \equiv \int \limits_{0}^{\infty} \,\frac{dx}{\sqrt{(x^2+a^2)(x^2+b^2)}} = \frac{\pi}{2 \cdot agm(a, b)}.
\end{displaymath}

To show that, we consider the integral
\begin{displaymath}
I(a_{N+1}, b_{N+1}) = \frac{1}{2} \int \limits_{-\infty}^{\infty} \,\frac{dx}{\sqrt{(x^2+a_{N+1}^2)(x^2+b_{N+1}^2)}}
\end{displaymath}
and perform the substitution
\begin{displaymath}
x = \frac{1}{2} \left(t-\frac{a_{N}b_{N}}{t}\right),\quad dx = \frac{1}{2} \left(1+\frac{a_{N}b_{N}}{t^2}\right) dt.
\end{displaymath}
The integral then reads
\begin{align*}
&I(a_{N+1}, b_{N+1})\\
&= \frac{1}{4} \int \,\frac{(1+a_{N}b_{N}/t^2)~dt}{\sqrt{\left[\frac{1}{4} \left(t-\frac{a_{N}b_{N}}{t}\right)^2+a_{N+1}^2\right] \left[\frac{1}{4} \left(t-\frac{a_{N}b_{N}}{t}\right)^2+b_{N+1}^2 \right]}}\\
&= \frac{1}{4} \int dt \,\frac{1+a_{N}b_{N}/t^2}{\sqrt{\left[ \frac{1}{4t^2} (t^2 + a_{N}^2)(t^2 + b_{N}^2)\right] \left[\frac{1}{4} t^2 \left(1+a_{N}b_{N}/t^2 \right) \right]}}\\
&= \int \,\frac{dt}{\sqrt{(t^2+a_{N}^2)(t^2+b_{N}^2)}}.
\end{align*}

For the integration limits we solve the quadratic equation for $t$
\begin{displaymath}
t_{1/2} = x \pm \sqrt{x^2+ab},
\end{displaymath}
which gives two equivalent results, since the integrand function is symmetric. Choosing $t_{1}$ we obtain
\begin{displaymath}
x \rightarrow \infty: t_{1} \rightarrow \infty,\quad x \rightarrow -\infty: t_{1} \rightarrow 0,
\end{displaymath}
thus the final result reads:
\begin{displaymath}
I(a_{N+1}, b_{N+1}) = \int \limits_{0}^{\infty} \,\frac{dt}{\sqrt{(t^2+a_{N}^2)(t^2+b_{N}^2)}} = I(a_{N}, b_{N}).
\end{displaymath}
This yields the general relation
\begin{displaymath}
I(a_{N+1}, b_{N+1}) = I \left(\frac{1}{2} \left(a_{N} + b_{N}\right), \sqrt{a_{N}b_{N}} \right).
\end{displaymath}
From this, it directly follows that $I(a_{N}, b_{N})$ is independent of $N$:
\begin{displaymath}
I(a_{0}, b_{0}) = I(a_{N}, b_{N}) = \lim_{n\rightarrow \infty} I(a_{n}, b_{n}) = I(c,c)
\end{displaymath}
with $c = agm(a,b)$.
Note that $I(c, c)$ is directly solvable:
\begin{displaymath}
I(c,c) = \int \limits_{0}^{\infty} \,\frac{dt}{\sqrt{(t^2+c^2)(t^2+c^2)}} = \int \limits_{0}^{\infty} \,\frac{dt}{t^2+c^2} = \frac{\pi}{2c} .
\end{displaymath}

The relation between the Gauss form for the complete elliptic integral of the first kind and the Legendre form $K(m)$ can be found by the substitution $\tan{\phi} = x/b$.

\end{document}